\documentclass[
notitlepage
,floatfix,
aps,
pra,
reprint,
superscriptaddress,
]{revtex4-2}
\usepackage[utf8]{inputenc}

\usepackage[caption=false]{subfig}
\usepackage{graphicx}
\usepackage{epstopdf}
\usepackage{array}
\usepackage{verbatim}
\usepackage{amsmath,amsfonts,amssymb,amscd,mathtools}
\usepackage{amsthm}
\usepackage{tabularx}
\usepackage{stmaryrd}
\usepackage{enumerate}
\usepackage{wasysym}
\usepackage{braket}
\usepackage{mathrsfs}
\usepackage{microtype}
\usepackage{hyperref}
\usepackage[dvipsnames]{xcolor}
\hypersetup{
    bookmarksnumbered=true, 
    unicode=false, 
    pdfstartview={FitH}, 
    pdftitle={}, 
    pdfauthor={}, 
    pdfsubject={}, 
    pdfcreator={}, 
    pdfproducer={}, 
    pdfkeywords={}, 
    pdfnewwindow=true, 
    colorlinks=true, 
    linkcolor=NavyBlue, 
    citecolor=NavyBlue, 
    filecolor=NavyBlue, 
    urlcolor=NavyBlue 
}
\usepackage{tikz}
\usepackage[lmargin=.7in,rmargin=.7in,tmargin=.7in,bmargin=1in]{geometry}

\usepackage{newtxtext,newtxmath}

\usepackage{titlesec} 

\usepackage{algorithm}
\usepackage{algorithmicx}
\usepackage{algpseudocode}



\usepackage{booktabs}

\theoremstyle{plain}
\newtheorem{thm}{Theorem}
\newtheorem{cor}[thm]{Corollary}
\newtheorem{lem}[thm]{Lemma}

\theoremstyle{definition}
\newtheorem{defn}[thm]{Definition}
\newtheorem{remark}[thm]{Remark}

\newcommand{\eq}[1]{(\hyperref[eq:#1]{\ref*{eq:#1}})}

\renewcommand{\sec}[1]{\hyperref[sec:#1]{Section~\ref*{sec:#1}}}
\newcommand{\thrm}[1]{\hyperref[thrm:#1]{Theorem~\ref*{thrm:#1}}}
\newcommand{\lemm}[1]{\hyperref[lemm:#1]{Lemma~\ref*{lemm:#1}}}
\newcommand{\prop}[1]{\hyperref[prop:#1]{Proposition~\ref*{prop:#1}}}
\newcommand{\corr}[1]{\hyperref[corr:#1]{Corollary~\ref*{corr:#1}}}
\newcommand{\fig}[1]{\hyperref[fig:#1]{~\ref*{fig:#1}}}
\newcommand{\deff}[1]{\hyperref[deff:#1]{~\ref*{deff:#1}}}

\newcommand{\mE}{\mathcal{E}}
\newcommand{\mN}{\mathcal{N}}
\newcommand{\mU}{\mathcal{U}}

\newcommand{\mD}{\mathcal{D}}
\newcommand{\mI}{\mathcal{I}}

\newcommand{\mM}{\mathcal{M}}
\newcommand{\mO}{\mathcal{O}}
\newcommand{\mB}{\mathcal{B}}

\newcommand{\mP}{\mathcal{P}}

\newcommand{\mZ}{\mathcal{Z}}

\newcommand{\mbI}{\mathbb{I}}

\newcommand{\mV}{\mathcal{V}}

\DeclareMathAlphabet{\matheu}{U}{eus}{m}{n}

\DeclareMathOperator{\tr}{tr}
\DeclareMathOperator{\Tr}{Tr}

\DeclareMathOperator{\sgn}{sgn}

\newcommand{\ketbra}[2]{|{#1}\rangle\!\langle{#2}|}

\newcommand{\ba}{\begin{eqnarray}}
\newcommand{\ea}{\end{eqnarray}}
\newcommand{\bann}{\begin{eqnarray*}}
\newcommand{\eann}{\end{eqnarray*}}
\newcommand{\bal}{\begin{equation}\begin{aligned}}
\newcommand{\eal}{\end{aligned}\end{equation}}
\newcommand{\dm}[1]{\ketbra{#1}{#1}}

\newcolumntype{L}[1]{>{\raggedright}p{#1}}
\newcolumntype{C}[1]{>{\centering}p{#1}}
\newcolumntype{R}[1]{>{\raggedleft}p{#1}}
\newcolumntype{D}{>{\centering\arraybackslash}X}

\renewcommand{\>}{\right\rangle}

\newcommand{\sbar}{\;\rule{0pt}{9.5pt}\right|\;}

\newcommand{\lset}{\left\{\left.}
\newcommand{\rset}{\right\}}

\DeclareMathOperator{\GHZ}{GHZ}

\begin{document}

\title{Fundamental limits of quantum error mitigation}

\author{Ryuji Takagi}
\email{ryuji.takagi@ntu.edu.sg}
\affiliation{Nanyang Quantum Hub, School of Physical and Mathematical Sciences, Nanyang Technological University, 637371, Singapore}

\author{Suguru Endo}
\email{suguru.endou.uc@hco.ntt.co.jp}
\affiliation{NTT Computer and Data Science Laboratories, NTT Corporation, Musashino, 180-8585, Tokyo, Japan}

\author{Shintaro Minagawa}
\email{minagawa.shintaro@nagoya-u.jp}
\affiliation{Graduate School of Informatics, Nagoya University, Chikusa-ku, 464-8601, Nagoya, Japan}

\author{Mile Gu}
\email{mgu@quantumcomplexity.org}
\affiliation{Nanyang Quantum Hub, School of Physical and Mathematical Sciences, Nanyang Technological University, 637371, Singapore}
\affiliation{Centre for Quantum Technologies, National University of Singapore, 3 Science Drive 2, 117543, Singapore}


\begin{abstract}
The inevitable accumulation of errors in near-future quantum devices represents a key obstacle in delivering practical quantum advantages, motivating the development of various quantum error-mitigation methods. Here, we derive fundamental bounds concerning how error-mitigation algorithms can reduce the computation error as a function of their sampling overhead. Our bounds place universal performance limits on a general error-mitigation protocol class. We use them to show  (1) that the sampling overhead that ensures a certain computational accuracy for mitigating local depolarizing noise in layered circuits scales exponentially with the circuit depth for general error-mitigation protocols and (2) the optimality of probabilistic error cancellation among a wide class of strategies in mitigating the local dephasing noise on an arbitrary number of qubits. Our results provide a means to identify when a given quantum error-mitigation strategy is optimal and when there is potential room for improvement.  

\end{abstract}

\maketitle


\begin{figure}[tbh!]
    \centering
    \includegraphics[width = 0.90\columnwidth]{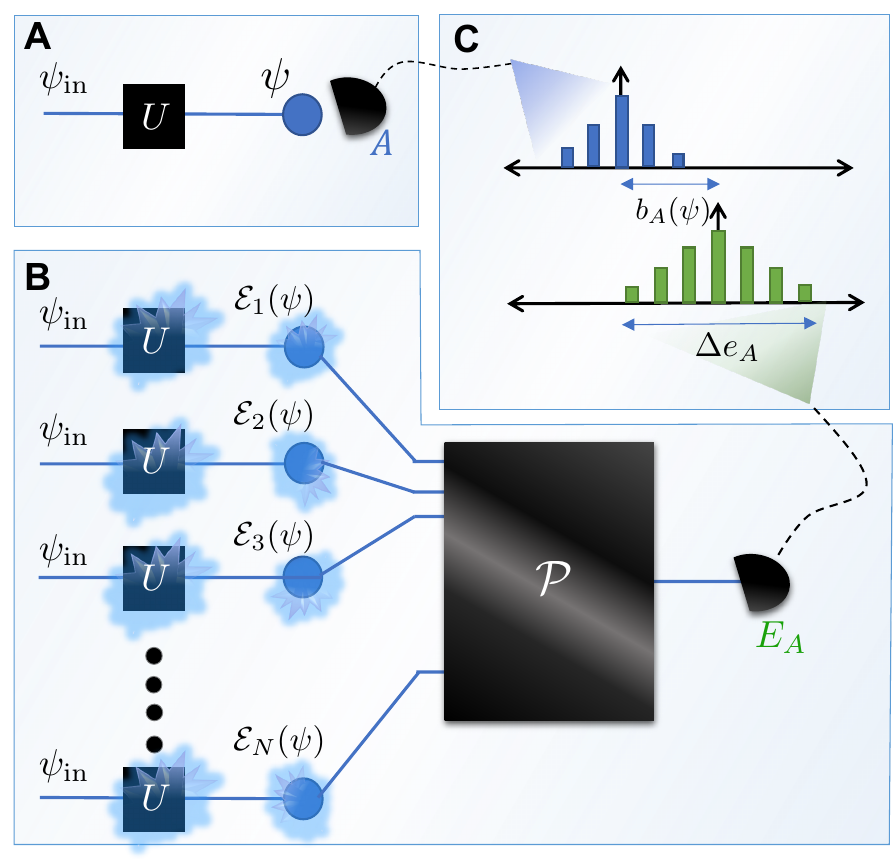}
     \caption{\textbf{Quantum error mitigation}. (A) A major goal of many near-term algorithms is to estimate the expectation value of some observable $A$, when acting on the output $\psi$ of some idealized computation $U$ applied to some input $\psi_{\rm in}$. (B) However, noise prevents the exact synthesis of $\psi$. Quantum error-mitigation protocols assist to estimate the true expectation value $
     \langle A \rangle = \Tr(A\psi)$ without using the adaptive quantum operations necessary in general error correction. This is done by (1) using available NISQ devices to synthesize $N$ distorted quantum states $\{\mathcal{E}_n(\psi)\}_{n=1}^N$ and (2) acting some physical process $\mathcal{P}$ on these distorted states to produce a random variable $E_A$ that approximates $A$.
     This procedure can then be repeated over $M$ rounds to draw $M$ samples of $E_A$, whose mean is used to estimate $ \langle A \rangle$. (C) We can characterize the efficacy of such protocol by (1) its spread $\Delta e_A$, the difference between maximum and minimum possible values of $E_A$ and (2) the bias $b_A(\psi) = \<E_A\> -  \langle A \rangle $. Here we derive ultimate lower bounds on $\Delta e_A$ for each given bias that no such error-mitigation protocol can exceed, as well as tighter bounds when $\mathcal{P}$ is restricted only to coherent interactions over $Q$ noisy devices at a time. This then tells us how many times $\mathcal{P}$ must be executed to estimate $\langle A \rangle$ within some desired accuracy and failure probability.}  
    \label{fig:framework}
\end{figure}

\section{Introduction}

Recent advances in quantum technologies have resulted in the availability of noisy intermediate-scale quantum (NISQ) devices, promising advantages of quantum information processing by controlling tens to hundreds of qubits~\cite{Preskill2018quantumcomputingin,Arute2019quantum}. However, inevitable noise remains a critical roadblock for their practical use; every gate has a chance of error, and their continuing accumulation will eventually destroy any potential quantum advantage. While quantum error correction enables in-principle means to suppress such error indefinitely, they involve measuring error syndromes and making adaptive corrections. In contrast, NISQ devices often cannot adaptively execute quantum operations.

This technological hurdle has motivated the study of quantum error mitigation, resulting in a diverse collection of alternative techniques (e.g., zero-error noise extrapolation~\cite{Temme2017error,Li2017efficient,giurgica2020digital,he2020zero,Kandala2019error,dumitrescu2018cloud}, probabilistic error cancellation~\cite{Buscemi2013twopoint,Buscemi2014twopoint,Temme2017error,Endo2018practical,song2019quantum,zhang2020error}, and virtual distillation~\cite{koczor2021exponential,huggins2021virtual,czarnik2021qubit,cai2021resource,huo2021dual,xiong2021quantum}). All share in common that they avoid adaptive operations. Instead, error-mitigation algorithms suppress errors by sampling available noisy devices many times and classically post-processing these measurement outcomes. Such techniques generally have drastically reduced technological requirements, providing potential near-term solutions for suppressing errors in other NISQ algorithms (e.g., variational algorithms for estimating the ground state energy in quantum chemistry~\cite{kandala_hardware-efficient_2017,McArdle2020quantum,cao2019quantum,mcardle2019error}).

The performance of these algorithms is typically analyzed on a case-by-case basis. 
While this is crucial for understanding the value of a particular methodology in a specific practical context, it leaves open a fundamental question: What is the ultimate potential of quantum error mitigation? The motivation to answer this question parallels the development of heat engines. There, Carnot's theorem allows us to understand the ultimate efficiency of all possible heat engines~\cite{Carnot1824reflections}, allowing us to know what is physically forbidden and enabling a universal means to understand what specific engines have the greatest room for potential improvement.

Here, we initiate a research program toward characterizing the ultimate limits of quantum error mitigation. We propose a framework to formally define error mitigation as any strategy that requires no adaptive quantum operations (see Fig.~\ref{fig:framework}).
We introduce maximum estimator spread as a universal benchmark for error-mitigation performance --- a quantity that tells us how many extra runs of a NISQ device guarantee that outputs are within some desired accuracy threshold.
We then derive fundamental lower bounds for this spread --- that no current or yet-undiscovered error-mitigation strategy can violate.
Our bounds are represented in terms of the reduction in the distinguishability of quantum states due to the noise effect, providing an operational understanding of the cost for error mitigation.

We then illustrate two immediate consequences of our general bounds. The first is in the context of mitigating local depolarizing noise in variational quantum circuits~\cite{kandala_hardware-efficient_2017,wang2021noiseinduced}. 
We show that the maximum estimator spread grows exponentially with circuit depth for the general error-mitigation protocol, confirming a suspicion that the well-known exponential growing estimation error observed in several existing error-mitigation techniques ~\cite{Temme2017error,Yuan2016simulating} is a consequence of the fundamental obstacle shared by the general error-mitigation strategies. 
Our second study shows that probabilistic error cancellation --- a prominent method of error mitigation --- minimizes the maximum estimator spread when mitigating local dephasing noise acting on an arbitrary number of qubits. These results showcase how our bounds can help rule out what error-mitigation performance targets are unphysical, and identify what methods are already near-optimal.

\section{Results}

\textbf{Framework} ---  Our framework begins by introducing a formal definition of error mitigation. Consider an ideal computation described by (1) application of some circuit $U$  to some input $\psi_{\rm in}$  (2) measurement of the output state $\psi$ in some arbitrary observable $A$ (See Fig.\ref{fig:framework}A). In realistic situations, however, there is noise, such that we have only access to NISQ devices capable of preparing some certain distorted states $\mE(\psi)$. The aim is then to retrieve desired output data specified by $\langle A \rangle = \Tr(A\psi)$. Here, we assume $-\mbI/2\leq A\leq \mbI/2$ without loss of generality. This is because any observable $O$ can be shifted and rescaled to some $A$ satisfying this condition, from which full information of $O$ can be recovered. For instance, if we are interested in a non-identity Pauli operator $P$, which has eigenvalues $\pm 1$, we instead consider an observable $A=P/2$. Note also that while $\psi$ is pure in many practically relevant instances, our analysis applies equally when $\psi$ is mixed.

We consider NISQ devices with no capacity to execute adaptive quantum operations. That is, they cannot enact different quantum operations conditioned on a measurement outcome. We then refer to an algorithm aimed to estimate $\langle A \rangle$ under such constrained devices as an error-mitigation strategy. Each error-mitigation strategy involves sampling NISQ devices configured in $N$ settings for some integer $N$. 
Denote the states generated by these configurations by $\mE_1(\psi),\dots,\mE_N(\psi)$, with effective noise channels $\{\mE_i\}_{i=1}^N$, where these effective noise channels can be different from each other in general. 
The effective noise channel is a non-adaptive operation that connects an ideal state to a distorted state and may be different from the actual noise channel that happens in the NISQ device. Nevertheless, one can always find such an effective noise channel given the descriptions of the actual noise channels and the idealized circuit $U$. 
The strategy then further describes some physical process $\mathcal{P}$ --- which is independent of either the input $\psi_{\rm in}$ or the ideal output $\psi$ --- that takes these distorted states as input and outputs some classical estimate random variable $E_A$ of $\Tr(A\psi)$ (See Fig.\ref{fig:framework}B). 
The aim is to generate $E_A$ such that its expected value $\langle E_A \rangle$ is close to $\Tr(A\psi)$. Each round of the protocol involves generating a sample of $E_A$. $M$ rounds of this procedure then enable us to generate $M$ samples of $E_A$, whose mean is used to estimate $\Tr(A\psi)$. 

\begin{figure}
    \centering
    \includegraphics[width=\columnwidth]{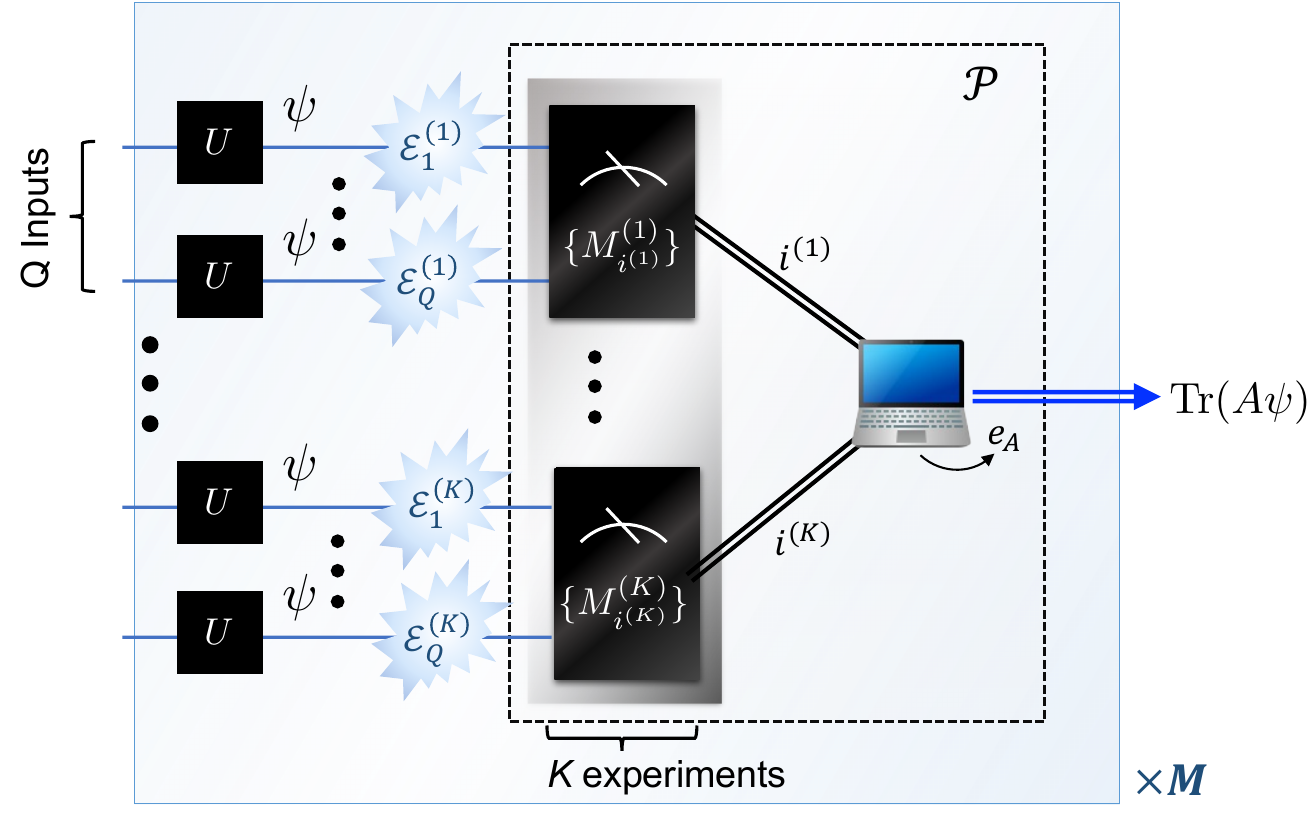}
    \caption{\textbf{Schematic of a $(Q,K)$-error mitigation protocol}. A $(Q,K)$-error mitigation protocol is motivated when practical considerations limit the maximum number of distorted states that our mitigation process $\mathcal{P}$ can coherently interact to $Q$. A general approach then divides these into $K=\lceil N/Q \rceil$ groups of size $Q$.  To estimate $\langle A \rangle$ of some ideal output state $\psi$, each round of  $(Q,K)$-mitigation involves first using available NISQ devices to generate $Q$ copies of each distorted states $\mE_q^{(k)}(\psi)$, for each of $k = 1, \ldots K$. These distorted states are then grouped together as inputs into $K$ experiments, where each group consists of a single copy of each $\mE_q^{(k)}(\psi)$. The $k^{\mathrm{th}}$ experiment then involves applying some general (possibly entangling) POVM $\{M_{i^{(k)}}^{(k)}\}$ on the $k^{\mathrm{th}}$ grouping, resulting in measurement outcome $i^{(k)}$. Classical computing is then deployed to produce an estimate $e_A(i^{(1)},\dots,i^{(K)})$ whose average after $M$ rounds of the above process is used to estimate $\mathrm{Tr}(A \psi)$. Note that there can be additional quantum operations before the POVM measurements $\{M_{i^{(k)}}^{(k)}\}$, but these can be absorbed into the description of the POVMs without loss of generality.}
    \label{fig:error_mitigation}
\end{figure}

Each error-mitigation strategy can then be entirely described by its choice of $\mathcal{P}$ and $\{\mE_i\}_{i=1}^N$. Our most fundamental bound pertain to all possible choices. However, we can often make these bounds tighter in situations where further practical limitations constrain how many distorted states $\mathcal{P}$ can coherently interact. Error mitigation protocols under such constraints typically select $N = KQ$ to a multiple of $Q$, such that the $N$ distorted states are divided into $K$ clusters, each containing $Q$ distorted states. We label these as $\{\mE_q^{(k)}(\psi)\}_{q=1, k=1}^{Q, K}$ for convenience. $\mathcal{P}$ is then constrained to represent (1) local measurement procedures $M^{(k)}$ that can coherently interact distorted states within the $k^{\mathrm{th}}$ cluster (i.e., $\{\mE_q^{(k)}(\psi)\}_{q=1}^{Q}$) to produce some classical interim outputs $i^{(k)}$ and (2) classical post-processing function $e_A$ that transform the interim outputs $\{i^{(k)}\}_{k=1}^K$ into a sample of $E_A$. 

\begin{figure}
    \centering
    \includegraphics[width=\columnwidth]{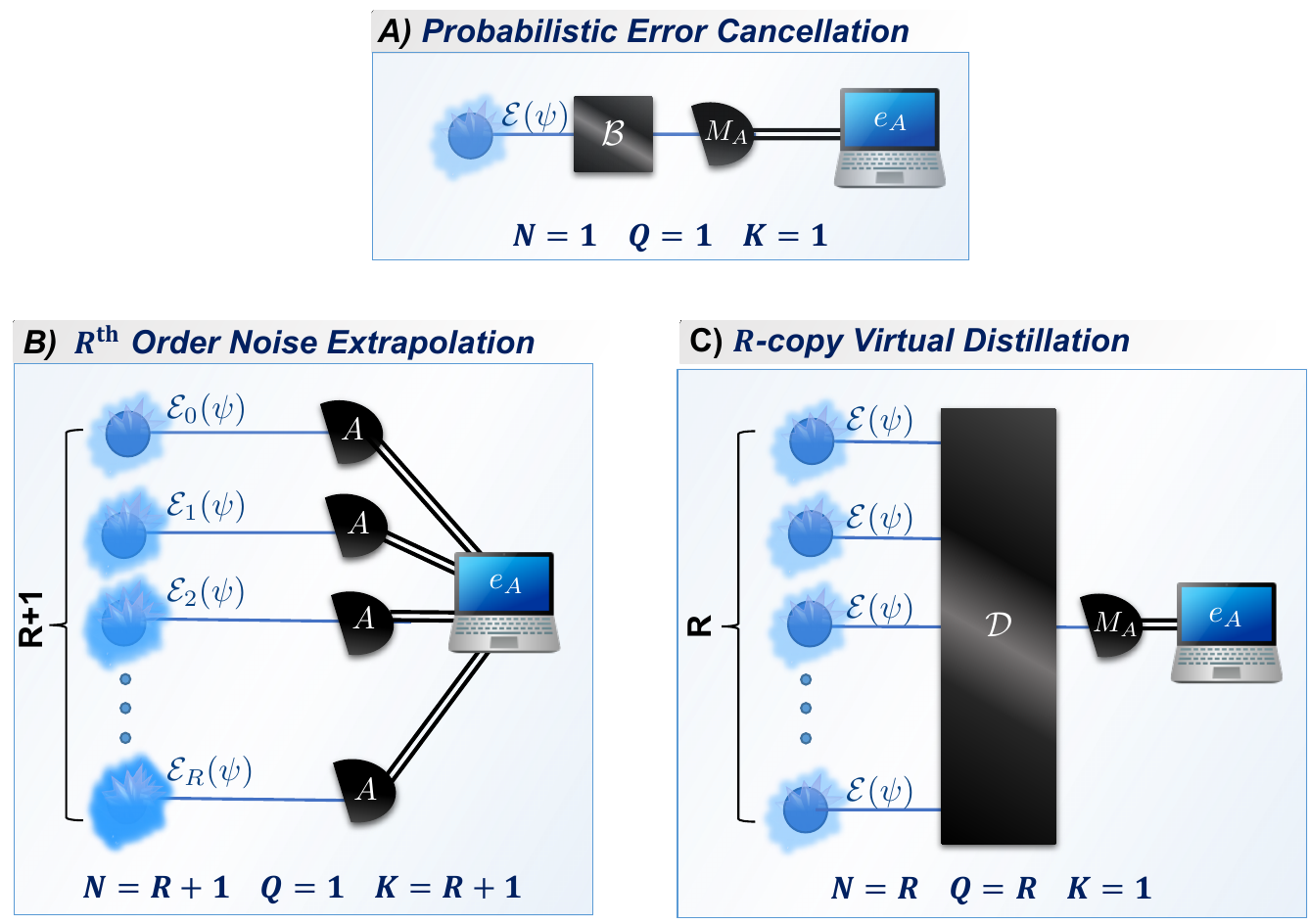}
    \caption{\textbf{Error-mitigation protocols}. Our framework encompasses all commonly used error-mitigation protocols, a sample of which we outline here. (A) Probabilistic error cancellation~\cite{Temme2017error} assumes we can only act a single coherent state each round, where it seeks to undo a given noise map $\mathcal{E}$ by applying a suitable stochastic operation $\mathcal{B}$. Thus it corresponds to the case of $Q=K=1$. (B) $R^{\mathrm{th}}$ order noise extrapolation assumes~\cite{Temme2017error,Li2017efficient} the capacity to synthesize $R+1$ NISQ devices whose outputs represent distortions of $\psi$ at various noise strengths. It then uses individual measurements of an observable $A$ on these distorted states to estimate the observable expectation value on the zero-noise limit. Thus it is an example where $Q=1$ and $K = R+1$. (C) Meanwhile, $R$-copy virtual distillation~\cite{koczor2021exponential,huggins2021virtual} involves running an available NISQ device $R$ times to synthesize $R$ copies of a distorted state $\mathcal{E}(\psi)$. Coherent interaction $\mD$ over these copies followed by a suitable measurement $M_A$ then enables improved estimation of $\langle A \rangle$. Thus it is an example where $K = 1$ and $Q = R$. In the main text and Methods, we provide a detailed account of each protocol and how it fits within our framework. }
    \label{fig:protocols}
\end{figure}

We name such a protocol as $(Q,K)$-error mitigation, and refer to the generation of each $i^{(k)}$ as an experiment. Each round of a $(Q,K)$-error mitigation protocol thus contains $K$ experiments on systems of up to $Q$ distorted states. 
We also summarize the above procedure in Fig.~\ref{fig:error_mitigation} and give a formal mathematical definition in Methods. Fig.~\ref{fig:protocols} and accompanying captions discuss how several prominent error-mitigation methods fit into this framework.

Several comments on our error-mitigation framework are in order.
We first note that, for a given set of noisy circuits that result in effective noise channels $\{\mE_i\}_{i=1}^N$, our framework assumes to apply an additional process $\mP$ after the noisy circuits and does not include processes within the initial noisy circuits.
Our framework thus excludes error correction, which employs adaptive processes integrated into noisy circuits.
This allows our framework to differentiate error mitigation from error correction and makes it useful to investigate the limitations imposed particularly on the former.

One might think that this would overly restrict the scope of error mitigation, which could also use some processes in noisy circuits. 
This can be avoided by considering that such processes are already integrated into the description of effective noise channels $\{\mE_i\}_{i=1}^N$. 
In other words, the effective noise channel can be considered as a map that connects an ideal state to a distorted state affected by not only a noise channel but non-adaptive processes accessible to a given near-term device; the error mitigation process $\mP$ is then an additional process that follows them.
This is manifested in the $R^{\mathrm{th}}$ order noise extrapolation in Fig.~\ref{fig:protocols}\,B, in which $R$ different noise levels realized on a near-term device are represented by the set $\{\mE_i\}_{i=1}^R$ of effective noise channels.

More broadly, taking appropriate effective noise channels allows our framework to include error-mitigation protocols that employ modified circuits. Namely, if $\{\mN_i\}_{i=1}^N$ are the noisy circuits that an error-mitigation protocol employs and $\mU$ is the ideal circuit, then such an error-mitigation strategy is encompassed in our framework with $\mE_i=\mN_i\circ\mU^\dagger$. This, for instance, includes the conventional strategy of probabilistic error cancellation applied to a noisy circuit, in which a probabilistic operation is applied after every noisy gate. 

We also remark that our framework leaves the freedom of how to choose the round number $M$ and the sample number $N=KQ$ per round for a given shot budget; if the total shot budget is $T$, one is free to choose any $N$ and $M$ such that $T=NM$.
As we describe shortly, our results in Theorem~\ref{thm:spread} and Corollary~\ref{cor:samples} are concerned with the number of rounds $M$, and they apply to any choice of shot allocation. 
However, our results become most informative by choosing as large $M$ (equivalently, as small $N$) as possible. 
The strategies in Fig.~\ref{fig:protocols} admit small $N$'s that do not scale with the total shot budget, representing examples for which our results give fruitful insights into their round number $M$.  
On the other hand, some strategies that employ highly nonlinear computation on the measurement outcomes (e.g., exponential noise extrapolation~\cite{Endo2018practical}, subspace expansion~\cite{McClean2017hybrid}) require a large $N$, in which case our results on the round number $M$ can have a large gap from the actual sampling cost.    

Our framework also allows one to assume some pre-knowledge prior to the error-mitigation process.
For instance, this includes the information about the underlying noise or some pre-computation that error-mitigation process can use in its strategy. 
The results in Theorem~\ref{thm:spread} and Corollary~\ref{cor:samples} then give information about the round number $M$ given such pre-knowledge.
Since the process of obtaining the pre-knowledge itself may be considered as a part of error-mitigation process, there are many possible divisions between the pre-computation and the error-mitigation process.
Our results apply to any choice of pre-knowledge, and this can be flexibly chosen depending on one's interest.
For instance, $R$-copy virtual distillation can be considered as a $(R,1)$-error mitigation (that is, $N=R$) as in Fig.~\ref{fig:protocols}\,C under the pre-knowledge of an eigenvalue of the noisy state, which is one of the settings discussed in Ref.~\cite{koczor2021exponential} (see also Methods). 
This pre-knowledge allows for a small choice of $N$, making the estimation of the round number $M$ by our method insightful. 
Another example includes the Clifford Data Regression~\cite{Czarnik2021errormitigation}, which can employ a linear regression based on a pattern learned from a training set. 
By considering the first learning step as the pre-computation, our results provide a meaningful bound for the sampling cost in the latter stage in which the output from the circuit of interest is compared to the model estimated from the training set.  

Up to the flexibility described above, our framework encompasses a broad class of error-mitigation strategies proposed so far~\cite{Temme2017error,Li2017efficient,Endo2018practical,McClean2017hybrid,Bonet2018lowcost,koczor2021exponential,huggins2021virtual,Bravyi2021mitigating,Yoshioka2021generalized,mcclean2020decoding,Czarnik2021errormitigation}.

\textbf{Quantifying performance} --- The performance of an error-mitigation protocol is determined by how well the random variable $E_A$ governing each estimate aligns with $\Tr(A\psi)$. We can characterize this by (1) its bias, representing how close $\langle E_A \rangle$ is to the ideal expectation value $\Tr(A\psi)$ and (2) its spread, representing the amount of intrinsic randomness within $E_A$.

A protocol's bias quantifies the absolute minimum error with which it can estimate $\mathrm{Tr}(A\psi)$, given no restrictions on how many rounds it can run (i.e., samples of $E_A$ it can draw). Mathematically, this is represented by the difference $b_A(\psi) =   \langle E_A \rangle - \Tr(A\psi)$. Since the error-mitigation strategy should work for an arbitrary state $\psi$ and observable $A$, we can introduce the maximum bias 
\bal
 b_{\max}\coloneqq \max_{-\mbI/2\leq A \leq \mbI/2}\max_\psi \,  |\langle E_A \rangle - \Tr(A\psi)|
\eal
to bound the bias of an error-mitigation protocol in estimating expectation values over all output states and observables of interest. Hereafter, we will also assume $b_{\max}\leq 1/2$, as this condition must be satisfied for any meaningful error-mitigation protocol. This is because a maximum bias of $1/2$ can always be achieved by the trivial `error-mitigation' protocol that outputs $e_A = 0$ regardless of $\psi$ or $A$.

Of course, having $b_{\max} = 0$ still does not guarantee an effective error-mitigation protocol. Each sample of $E_A$ will also deviate from $\Tr(A\psi)$ due to intrinsic random error. The greater this randomness, the more samples we need from $E_A$ to ensure that the mean of our samples is a reliable estimate of its true expectation value $\langle E_A \rangle$. The relation is formalized by Hoeffding's inequality~\cite{hoeffding1963probability}. 
Namely, suppose $\{x_i\}_{i=1}^{M}$ are $M$ samples of a random variable $X$ with $x_i \in [a,b]$, the number $M$ of samples that ensures an estimation error $|\<X\>-\sum_ix_i/M|<\delta$ with probability $1-\varepsilon$ is given by $\frac{|a-b|^2}{2\delta^2}\log(2/\varepsilon)\propto |a-b|^2$. 
In our context, the latter quantity corresponds to the maximum spread in the outcomes of estimator function $e_A$ defined by
\bal
  \Delta e_{\max} \coloneqq  \max_{-\mbI/2\leq A \leq \mbI/2} \Delta e_{A},  
  \label{eq:spread definition}
 \eal
where $\Delta e_{A}$ is the difference between the maximum and minimum possible values that $E_A$ can take, i.e., $\Delta e_{A}\coloneqq e_{A,\max}-e_{A,\min}$ where $e_{A,\max} \coloneqq \max_{i^{(1)}\dots i^{(K)}} e_A(i^{(1)}\dots i^{(K)})$ and $e_{A,\min} \coloneqq \min_{i^{(1)}\dots i^{(K)}}e_A(i^{(1)}\dots i^{(K)})$.

$\Delta e_{\max}$ thus directly relates to the sampling cost of an error-mitigation protocol. Given an error-mitigation protocol whose estimates have maximum spread $\Delta e_{\max}$, it uses sample $E_A$ of order $\mO(\Delta e_{\max}^2\log(1/\varepsilon)/\delta^2)$ times to ensure that its estimate of $\langle E_A \rangle$ has accuracy $\delta$ and failure rate $\varepsilon$. Therefore, we may think of $\Delta e_{\max}$ as a measure of computational cost or feasibility. 
Its exponential scaling with respect to the circuit depth, for example, would imply eventual intractability in mitigating associated errors in a class of non-shallow circuits.

We note that if the variance of $E_A$ happens to be small, the actual sampling cost required to achieve the accuracy $\delta$ and failure rate $\varepsilon$ can be smaller than the estimate based on the maximum spread. In this sense, $\Delta e_{\max}$ quantifies the round number $M$ that one would practically use in the worst-case scenario. 
However, knowing the variance of $E_A$ beforehand is a formidable task in general, and the worst-case estimate gives a useful benchmark to assess the feasibility of a given error-mitigation strategy in such situations.


\textbf{Fundamental limits} ---
Our main contribution is to establish a universal lower bound on $\Delta e_{\max}$. Our bound then determines the number of times an error-mitigation method samples $E_A$ (and thus the number of times we invoke a NISQ device) to estimate $A$ within some tolerable error.  

To state the bound formally, we utilize measures of state distinguishability. Consider the scenario where Alice prepares a quantum state in either $\rho$ and $\sigma$ and challenges Bob to guess which is prepared. The trace distance $D_{\rm tr}(\rho,\sigma) = \frac{1}{2}\|\rho-\sigma\|_1$ (where $\|\cdot\|_1$ is the trace norm) then represents the quantity such that Bob's optimal probability of guessing correctly is $\frac{1}{2}(1 +  D_{\rm tr}(\rho,\sigma))$. When $\rho$ and $\sigma$ describe states on $K$-partite systems $S_1\otimes\dots\otimes S_K$, we can also consider the setting in which Bob is constrained to local measurements, resulting in the optimal guessing probability $\frac{1}{2}(1+D_{\rm LM}(\rho,\sigma))$ where $D_{\rm LM}$ is the local distinguishability measure~\cite{matthews_distinguishability_2009} (see also Methods).
In our setting, we identify each local subsystem $S_k$ with a system corresponding to the $k^{\mathrm{th}}$ experiment in Fig.~\ref{fig:error_mitigation}. We are then in a position to state our main result:

\begin{thm}\label{thm:spread} Consider an arbitrary $(Q,K)$-mitigation protocol with maximum bias $b_{\max}$. Then, its maximum spread $\Delta e_{\max}$ is lower bounded by
\bal 
 \Delta e_{\max}&\geq\max_{\substack{\psi,\phi}}\frac{D_{\rm tr}(\psi,\phi)-2b_{\max}}{D_{\rm LM}\left(\tilde\psi_Q^{(K)},\tilde\phi_Q^{(K)}\right)} 
\label{eq:spread bound}
\eal
where $\tilde\psi_Q^{(K)}\coloneqq\otimes_{k=1}^{K}\otimes_{q=1}^{Q}\left[\mE_q^{(k)}(\psi)\right]$ and $\tilde\phi_Q^{(K)}\coloneqq\otimes_{k=1}^{K}\otimes_{q=1}^{Q}\left[\mE_q^{(k)}(\phi)\right]$ are distorted states corresponding to the $QK$ copies of some ideal outputs $\psi$ and $\phi$, and $\mE_q^{(k)}$ is the effective noise channel for the $q^{\mathrm{th}}$ input in the $k^{{\mathrm{th}}}$ experiment.
\end{thm}

Combining this with Hoeffding's inequality leads to the following bound on the sampling cost.

\begin{cor}\label{cor:samples} Consider an arbitrary $(Q,K)$-mitigation protocol with maximum bias $b_{\max}$. 
Then, an estimation error of $b_{\max}+\delta$ is realized with probability  $1-\varepsilon$ when the number of samples $M$ satisfies
\bal 
 M&\geq \frac{\Delta e_{\max}^2\log(2/\varepsilon)}{2\delta^2}\\
 &\geq\left[\max_{\substack{\psi,\phi}}\frac{D_{\rm tr}(\psi,\phi)-2b_{\max}}{D_{\rm LM}\left(\tilde\psi_Q^{(K)},\tilde\phi_Q^{(K)}\right)}\right]^2 \frac{\log(2/\varepsilon)}{2\delta^2} 
\label{eq:samples bound}
\eal
where $\tilde\psi_Q^{(K)}\coloneqq\otimes_{k=1}^{K}\otimes_{q=1}^{Q}\left[\mE_q^{(k)}(\psi)\right]$ and $\tilde\phi_Q^{(K)}\coloneqq\otimes_{k=1}^{K}\otimes_{q=1}^{Q}\left[\mE_q^{(k)}(\phi)\right]$.
\end{cor}

\noindent Theorem~\ref{thm:spread} and Corollary~\ref{cor:samples} offer two qualitative insights. The first is the potential trade-off between sampling cost and systematic error --- we may reduce the sampling cost by increasing tolerance for bias. The second is a direct relation between sampling cost and distinguishability --- the more a noise source degrades distinguishability between states, the more costly the error is to mitigate. 

The intuition behind this relation rests on the observation that the error-mitigation process is a quantum channel. 
Thus, any error-mitigation procedure must obey data-processing inequalities for distinguishability. On the other hand, error mitigation aims to improve our ability to estimate expectation values of various observables, which would enhance our ability to distinguish between noisy states. The combination of these observations then implies that distinguishability places a fundamental constraint on required sampling costs to mitigate error. For details of the associated proof, see Methods.

Observe that our bound involves the local distinguishability $D_{\textrm{LM}}(\rho,\sigma)$ rather than the standard trace distance $D_{\rm tr}(\rho,\sigma)$. This is due to the constraints we placed of $\mathcal{P}$ that limits it to coherently interacting the outputs of a finite number of NISQ devices --- reflecting the hybrid nature of quantum error mitigation utilizing quantum and classical resources in tandem. Notably, these quantities coincide for the most powerful NISQ devices (the ones allowing coherent interactions between all $N$ noisy initial states). This case then corresponds to the most fundamental bound 
\bal 
 \Delta e_{\max}&\geq \max_{\substack{\psi,\phi}}\frac{D_{\rm tr}(\psi,\phi)-2b_{\max}}{D_{\rm tr}\left(\tilde\psi_Q^{(K)},\tilde\phi_Q^{(K)}\right)},
\label{eq:spread bound2}
\eal
which represents the ultimate performance limits of all $(Q,K)$ error-mitigation protocols that coherently operate on $N = QK$ distorted states each round.

We also remark that our framework can give tighter bounds when available error-mitigation methods involve specific states and observables (see Eq.~\eqref{eq:spread fixed observable}).

\textbf{Alternative bounds} --- While the bounds derived above in terms of distinguishability have a clear operational meaning, its evaluation in realistic settings can face two significant hurdles. (1) It involves evaluating the  distinguishability between two quantum states whose dimensions scale exponentially with $KQ$, making its evaluation costly for protocols that require many NISQ samples per round. (2) It requires that we have tomographic knowledge of the effective noise channels $\mE_q^{(k)}$. 

One potential means around this is to identify bounds on the distinguishability measures that alleviate such hurdles.
For example, since $D_{\rm tr}(\rho,\sigma)\leq \sqrt{1-F(\rho,\sigma)}$ for any pair of states $\rho$ and $\sigma$ where $F(\rho,\sigma)\coloneqq \left(\Tr\sqrt{\sigma^{1/2}\rho\sigma^{1/2}}\right)^2$ is the (squared) fidelity~\cite{Fuchs1999cryptographic}, this, together with Eq.~\eqref{eq:spread bound2}, implies
\bal 
 \Delta e_{\max}&\geq \max_{\substack{\psi,\phi}}\frac{D_{\rm tr}(\psi,\phi)-2b_{\max}}{\sqrt{1-\prod_{q=1}^Q\prod_{k=1}^K F\left(\mE_q^{(k)}(\psi),\mE_q^{(k)}(\phi)\right)}}.
\label{eq:fidelity bound}
\eal
This form only involves the computation of the trace distance and fidelity of single-copy states, both of which can be computed by semidefinite programming~\cite{Watrous2018theory}.

Meanwhile, the need for tomographic knowledge of $\mE_q^{(k)}$ can be mitigated by using subfidelity~\cite{Miszczak2008sub}
\bal
E(\rho,\sigma)\coloneqq \Tr(\rho\sigma)+\sqrt{2\left[\left\{\Tr(\rho\sigma)\right\}^2-\Tr(\rho\sigma\rho\sigma)\right]}.
\label{eq:subfidelity}
\eal
The subfidelity bounds $F(\rho,\sigma)$ from below, and thus also lower bounds the maximum spread:
\bal 
 \Delta e_{\max}&\geq \max_{\substack{\psi,\phi}}\frac{D_{\rm tr}(\psi,\phi)-2b_{\max}}{\sqrt{1-\prod_{q=1}^Q\prod_{k=1}^K E\left(\mE_q^{(k)}(\psi),\mE_q^{(k)}(\phi)\right)}}.
\label{eq:subfidelity bound}
\eal
subfidelity between two unknown states can be measured by a quantum computer using a circuit of constant depth~\cite{Ekert2002direct,Bacon2006efficient} (see also Methods). This obviates the need for tomographical data, while its low depth means that the noise in this process is typically much smaller than the noise in our circuits of interest. We remark that, instead of using the subfidelity, one could use an alternative quantity that lower bounds the fidelity that can be estimated by NISQ devices, e.g., truncated fidelity~\cite{Cerezo2020variational}. 
Such techniques could enable benchmarking protocols that allow us to rule out a candidate NISQ device should our bounds suggest their error profiles are too adverse to support any viable means of error mitigation.

In addition, the maximum in the right-hand sides of \eqref{eq:fidelity bound} and \eqref{eq:subfidelity bound} do not need to be evaluated exactly; any choice of states $\psi$ and $\phi$ provides a valid lower bound for the maximum spread.  
While these alternative bounds may not be as tight, they still serve as universal lower bounds that can put non-trivial constraints on the error-mitigation performance (see Remark~2 in Supplementary Note~1 and Supplementary Note~3).

\textbf{Error-mitigating layered circuits} ---
Quantitatively, the above bounds enable us to determine the ultimate performance limits of error mitigation given a particular set of imperfect quantum devices specified by error channels $\{\mE_q^{(k)}\}$. 
We now illustrate how this enables the identification of sampling overheads when performing error mitigation on a common class of NISQ algorithms --- layered circuits used extensively in variational quantum eigensolvers~\cite{peruzzo2014variational}. 
Variational algorithms typically assume a quantum circuit consisting of multiple layers of unitary gates $\{U_l\}_{l=1}^{L}$ acting on an $n$-qubit system. 
Indeed, as designed with NISQ applications in mind, they are key candidates for benchmarking of error-mitigation protocols~\cite{Kandala2019error,Kim2021scalable,Sagastizabal2019experimental}. 

\begin{figure}
    \centering
    \includegraphics[width=\columnwidth]{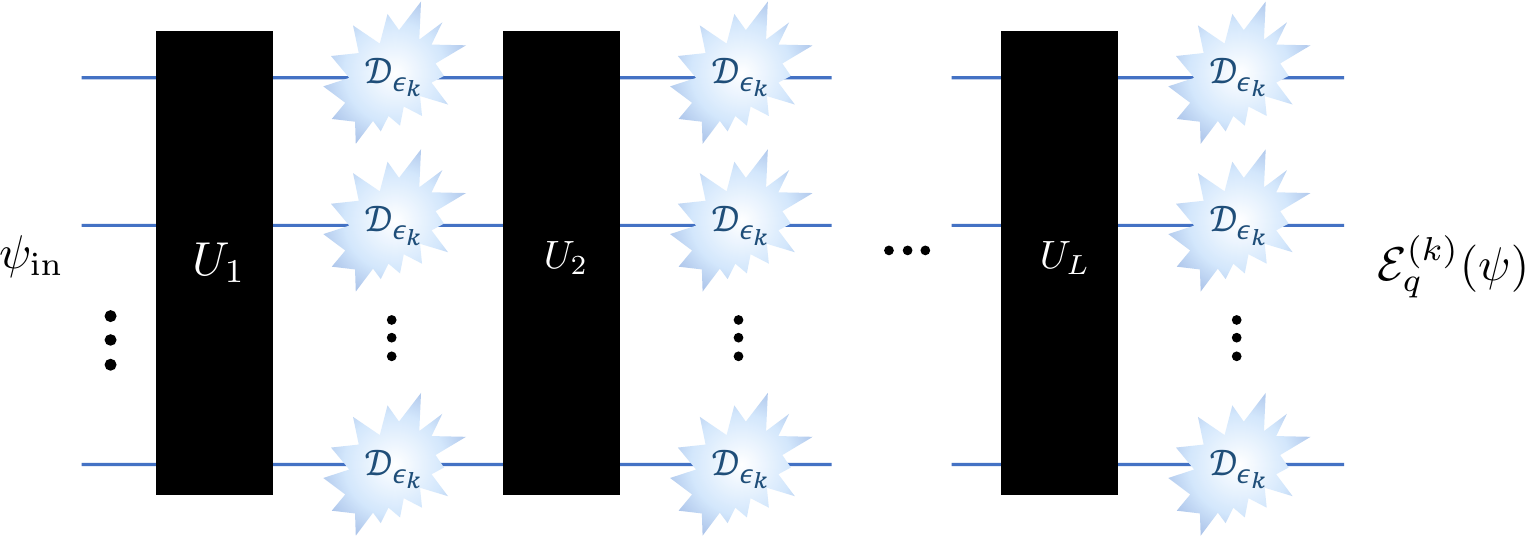}
    \caption{\textbf{Noise mitigation in layered circuits}. Layered circuits are used extensively in variational algorithms for NISQ devices. They involve repeated layers of gates, each consisting of some unitary $U_l$. A standard noise model for such circuits involves the action of local depolarizing noise $\mD_{\epsilon}$ on each qubit during each layer of the circuit. 
    The $k^{\mathrm{th}}$ experiment in a general $(Q,K)$-protocol involves running this circuit $Q$ times to produce a distorted state $\otimes_{q=1}^Q\mE_{q}^{(k)}(\psi)$ with some noise strength $\epsilon_k$ --- which possibly varies over different experiments.   
    The protocol then measures each $\otimes_{q=1}^Q\mE_{q}^{(k)}(\psi)$ for $k=1,\dots,K$ and outputs an estimate $E_A$
through classical post-processing of the measurements results.}
    \label{fig:noisy_layered}
\end{figure}

In particular, consider a local depolarizing noise~\cite{MullerHermes2016relative,wang2021noiseinduced}, in which the depolarizing channel $\mD_{\epsilon}(\rho)\coloneqq (1-\epsilon)\rho+\epsilon\mbI/2$ acts on each qubit.
A general approach to mitigate this error is to employ a $(Q,K)$-mitigation protocol for some $Q$ and $K$, in which the $k^{\mathrm{th}}$ experiment involves depolarizing noise with noise strength $\epsilon_k$ (Fig.~\ref{fig:noisy_layered}).

Taking $U=U_L\cdots U_2\,U_1$ in Fig.~\ref{fig:error_mitigation} and applying Theorem~\ref{thm:spread} to this setting, we obtain the following bound (See Supplementary Note~1 for the proof).

\begin{thm}\label{thm:layer}
For an arbitrary $(Q,K)$-error mitigation with maximum bias $b_{\max}$ applied to $n$-qubit circuits with $L$-layer unitaries under local depolarizing noise, the maximum spread is lower bounded as   
\bal
 \Delta e_{\max}\geq \frac{1-2b_{\max}}{\sqrt{2\ln 2} \sqrt{nQ}\, K}\left(\frac{1}{1-\epsilon_{\min}}\right)^{L},
\eal
where $\epsilon_{\min}\coloneqq\min_k \epsilon_k$ is the minimum noise strength among $K$ experiments. 
\end{thm}

Theorem~\ref{thm:layer} suggests that error-mitigation strategies encompassed in our framework will use exponentially many samples with respect to the circuit depth $L$. 
This validates our intuition that information should quickly get degraded due to the sequential noise effects, incurring exponential overhead to remove the accumulated noise effect. 

We also remark that, although we here focus on the exponential growth of the maximum spread with respect to the circuit depth $L$ for the sake of generality, one can expect that the maximum spread grows exponentially with the total gate number $nQKL$ rather than just the layer number $L$ in many practical cases.


\textbf{Protocol benchmarking} --- Theorems~\ref{thm:spread}~and~\ref{thm:layer} place strategy-independent bounds on the maximum spread for each $Q$ and $K$ and available noise channels $\mathcal{E}_q^{(k)}$, enabling us to identify the ultimate potential of error mitigation under various noise settings and operational constraints. Comparing this limit with that achieved by specific known methods of error mitigation then provides a valuable benchmark, helping us assess their optimality and quantify the potential room for improvement. We illustrate this here by considering probabilistic error cancellation~\cite{Temme2017error}, while we discuss how our framework can be applied to other prominent error-mitigation protocols in Methods. 

Probabilistic error cancellation is an error-mitigation protocol that produces an estimate of $\Tr(A\psi)$ using a distorted state $\mE(\psi)$ each round (see Fig \ref{fig:protocols}A).
It then fulfills the criteria of being a $(1,1)$-protocol, i.e., $Q=K=1$. 
Here, we assume that the description of the noise channels is given as pre-knowledge, in which case the estimator becomes unbiased, i.e., $b_{\max}=0$.
Probabilistic error cancellation operates by identifying a complete basis of processes $\{\mathcal{B}_j\}_j$ such that $\mE^{-1}=\sum_{j} c_{j} \mathcal{B}_{j}$ for some set of real (but possibly negative) numbers $\{c_j\}_j$. 
Setting 
$\gamma\coloneqq \sum_j|c_j|$, the protocol then (1) applies $\mB_j$ to the noisy state $\mE(\psi)$ with probability $p_j = |c_j|/\gamma$, (2) measures $A$ to get outcome $a_j$, and (3) multiplies each outcome by $\gamma\, \mathrm{sgn}(c_j)$ and takes the average.

In the context of our framework, we can introduce a quantum operation $\mathcal{B}$ that represents first initializing a classical register to a state $j$ with probability $p_j$ and applying $\mathcal{B}_{j}$ to $\mE(\psi)$ conditioned on $j$. 
Meanwhile, $M_A$ represents an $A$-measurement of the resulting quantum system combined with a measurement of the register, resulting in the outcome pair $(a_j, j)$. 
Taking $e_A^{\rm PEC}\left((a_j,j)\right) = \gamma\sgn(c_j)a_j$, we see that the maximum spread of this estimator is given by
\bal
 \Delta e_{\max}^{\rm PEC}=\gamma,
\eal
a well-studied quantity that is already associated with the sampling overhead of probabilistic error cancellation~\cite{Temme2017error}. 
 
The optimal sampling cost $\gamma_{\rm opt}$ is then achieved by minimizing such $\gamma$ over all feasible $\{\mathcal{B}_j\}_j$~\cite{takagi2020optimal}. 
Once computed for a specific noise channel $\mathcal{E}$, we can compare it to the lower bounds in Theorem~\ref{thm:spread} to determine if there is possible room for improvement.  

Let us now consider local dephasing noise on an $n$-qubit system, where the dephasing noise $\mZ_\epsilon(\rho)\coloneqq (1-\epsilon)\rho + \epsilon Z\rho Z$ acts on each qubit. 
We find that the optimal cost is obtained as
\bal
 \gamma_{\rm opt} = \Delta e_{\max}^{\rm PEC}= \frac{1}{(1-2\epsilon)^n}.
 \label{eq:PEC local dephasing achievable}
\eal
This can be compared to the bound for $\Delta e_{\rm max}$ from Theorem~\ref{thm:spread} that applies to every mitigation protocol with $Q=K=1$. Note that, since $K=1$, $D_{\rm LM}=D_{\tr}$. 
We then get 
\bal
 \max_{\psi,\phi}\frac{D_{\tr}(\psi,\phi)}{D_{\tr}(\mZ_\epsilon(\psi),\mZ_\epsilon(\phi))}&\geq \frac{1}{(1-2\epsilon)^n}.
 \label{eq:PEC local dephasing lower bound}
\eal
Detailed computation to obtain \eqref{eq:PEC local dephasing achievable} and \eqref{eq:PEC local dephasing lower bound} can be found in Supplementary Note~2. 
Remarkably, the two quantities --- the maximum spread for the probabilistic error cancellation and the lower bound for arbitrary unbiased mitigation strategies with $Q=K=1$ --- exactly coincide.   
This shows that probabilistic error cancellation achieves the ultimate performance limit of unbiased $(1,1)$-protocols for correcting local dephasing noise for an arbitrary qubit number $n$.

We can also consider the $d$-dimensional depolarizing noise $\mD_\epsilon^d(\rho)=(1-\epsilon)\rho + \epsilon\mbI/d$.
The bound from Theorem~\ref{thm:spread} for this noise is obtained as
\bal
 \max_{\psi,\phi}\frac{D_{\tr}(\psi,\phi)}{D_{\tr}(\mD_\epsilon^d(\psi),\mD_\epsilon^d(\phi))}=\frac{1}{1-\epsilon},
 \label{eq:PEC global depolarizing lower bound}
\eal 
which is slightly lower than $\Delta e_{\max}^{\rm PEC}=\frac{1+(1-2/d^2)\epsilon}{1-\epsilon}$~\cite{takagi2020optimal,Jiang2020physical,Regula2021operational}, with difference being $O(\epsilon)$.
This suggests that probabilistic error cancellation is nearly optimal for this noise model, while still leaving the possibility for a better protocol to exist.

\begin{figure}
    \centering
    \includegraphics[width=\columnwidth]{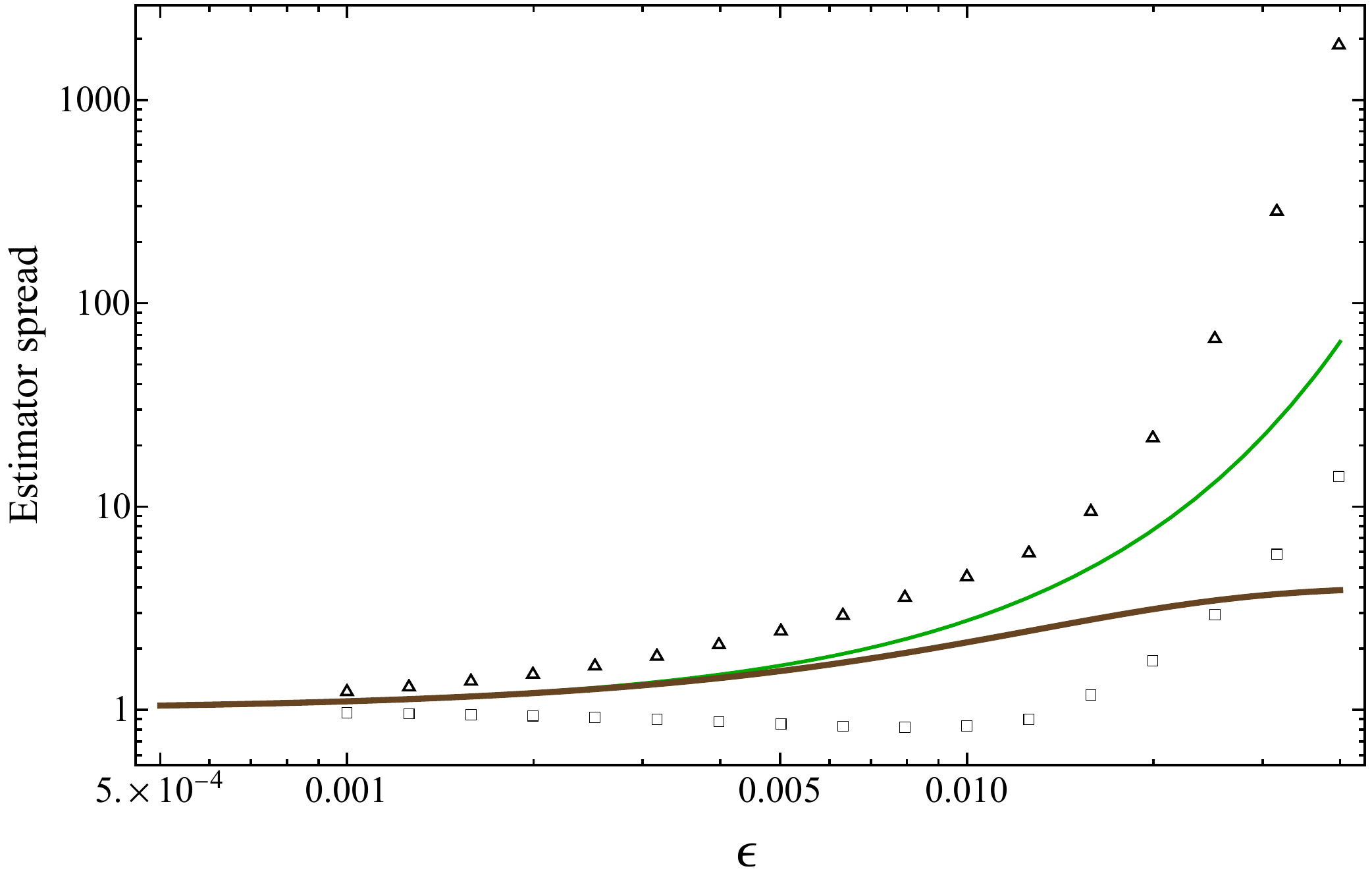}
    \caption{The estimator spreads to mitigate local dephasing noise on a 50-qubit system. Solid green curve: $\Delta e_{\max}$ for probabilistic error cancellation and the lower bound for unbiased $(1,1)$-mitigation protocols, which coincide as explained in the main text. Brown curve: $\Delta e_A$ with $A=\frac{1}{2}\otimes_{i=1}^n X_i$ for 2-copy virtual distillation with GHZ state inputs and the lower bound for $(2,1)$-mitigation protocols with the same bias, which coincide as explained in Supplementary Note~2. Triangles and rectangles: $\Delta e_A$ with $A=\frac{1}{2}\otimes_{i=1}^n X_i$ for $11^{\mathrm{th}}$ order noise extrapolation with GHZ state inputs (triangles) and a lower bound for $(1,12)$-mitigation protocols with the same bias (rectangles).}
    \label{fig:local_dephasing_all}
\end{figure}

\begin{figure}
    \centering
    \includegraphics[width=\columnwidth]{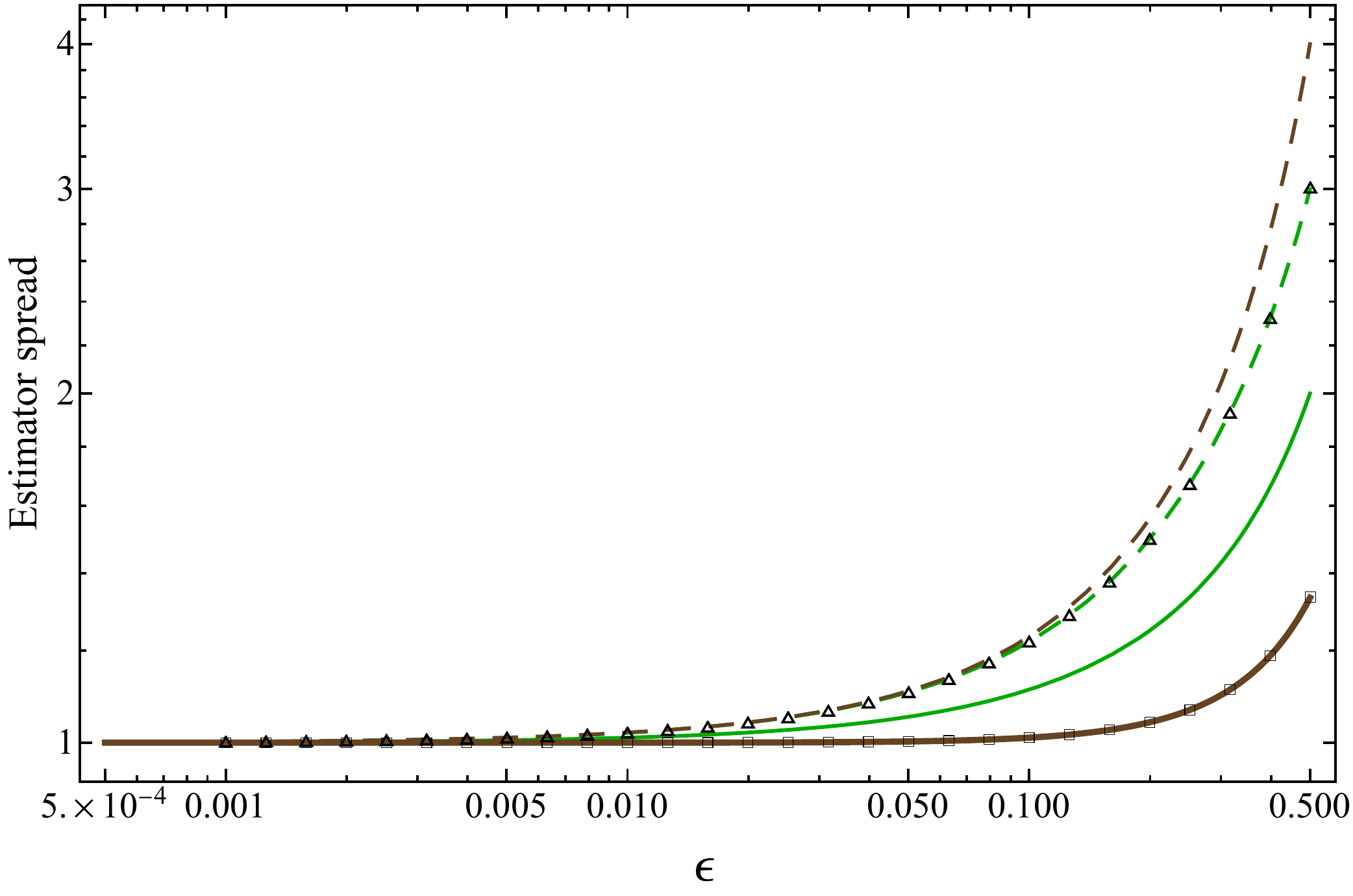}
    \caption{
    The estimator spreads to mitigate global depolarizing noise on a 50-qubit system. Green curves: $\Delta e_{\max}$ for probabilistic error cancellation (dashed) and the lower bound for unbiased $(1,1)$-mitigation protocols (solid). Brown curves: $\Delta e_A$ with $A=\frac{1}{2}\otimes_{i=1}^n X_i$ for 2-copy virtual distillation with GHZ state inputs (dashed) and the lower bound for $(2,1)$-mitigation protocols with the same bias (solid). Triangles and rectangles: $\Delta e_A$ with $A=\frac{1}{2}\otimes_{i=1}^n X_i$ for $1^{\mathrm{st}}$ order noise extrapolation (triangles) and a lower bound for $(1,2)$-mitigation protocols with the same bias (rectangles).}
    \label{fig:global_dep_all}
\end{figure}

We can also apply similar techniques to study the performance of other prominent error-mitigation protocols. Here, we plot the estimator spread for probabilistic error cancellation, virtual distillation, and noise extrapolation, and their corresponding lower bounds for local dephasing noise (Fig.~\ref{fig:local_dephasing_all}) and global depolarizing noise (Fig.~\ref{fig:global_dep_all}).
We note that, for virtual distillation and extrapolation, we evaluated \eqref{eq:spread fixed observable} that allows us to bound $\Delta e_A$ in \eqref{eq:spread definition} with a specific observable $A$ of interest.
We provide details for the evaluation of these values in Supplementary Note~2. We can observe that both protocols perform near-optimal limits at the low-error regime. At the high-error regime, their performance can diverge significantly from our lower bounds depending on underlying noise models and mitigation strategies.
We emphasize that such divergences are expected because of the high generality of our lower bounds. 
Narrowing the gaps between the fundamental lower bounds and achievable maximum spread, e.g., finding more examples such as probabilistic error cancellation for local dephasing noise, will be a natural direction for future work.

\section{Discussion}

Our work aimed to identify the ultimate performance limits of quantum error mitigation --- a large class of techniques designed to estimate the outputs of ideal quantum circuits by post-processing measurement data from imperfect counterparts. This involved identifying a universal performance measure --- applicable to any such error-mitigation protocols --- that captures how many extra executions of available NISQ devices the protocol uses to ensure that its estimates are sufficiently close with some required probability of success. We then derived ultimate performance limits that pertain to all such error mitigation methods. The significance of our bounds parallels that of various fundamental converse bounds in quantum information (e.g., quantum communication~\cite{Bennett1999entanglement,Pirandola2017fundamental,Berta2013entanglement} and thermodynamics~\cite{Landauer1961irreversibility,Brandao2015second,Gour2018quantum}), representing the ultimate performance limits that quantum error-mitigation protocols can never surpass. 
Our bounds particularly demonstrate that probabilistic error cancellation is optimal in the maximum spread to mitigate local dephasing noise among all unbiased error-mitigation protocols that involve no coherent interactions between multiple copies of distorted states, and imply that the exponential growth in the maximum spread on mitigating noise in layered circuits is an unavoidable feature shared by the general error-mitigation protocols.

We note that our performance bounds have focused on the scaling of $M$, representing how many rounds an error-mitigation protocol should be run to get a reliable estimate of some observable $\langle A \rangle$. 
Although this analysis is sufficient for many present methods of error mitigation, it is possible to also improve estimates of $\langle A \rangle$ by scaling the number of distorted outputs we process in a single round (e.g., extrapolation~\cite{Endo2018practical} and subspace expansion~\cite{McClean2017hybrid}). 
While our framework in Fig.~\ref{fig:framework} encompasses such methodologies --- and as such all bounds on estimation error apply --- full understanding of the performance of such protocols would involve further investigation on how estimation error scales with respect to $N$ or $K$. This then presents a natural direction for future research.

Our results also offer potential insights into several related fields. Non-Markovian dynamics have shown promise in decreasing sampling costs in error mitigation~\cite{hakoshima2021relationship}. Since non-Markovianity is known to be deeply related to the trace distance~\cite{breuer2016colloquium}, our newly established relations between trace distance and quantum error mitigation hint at promising relations between the two fields. 
The second direction is to relate our general framework of quantum error mitigation to the established theory of quantum error correction. 
Quantum error correction concerns algorithms that prevent degrading the trace distance between suitably encoded logical states, while our results indicate that less reduction in trace distance can enable smaller error mitigation costs. 
Thus, our work provides a toolkit for identifying fundamental bounds in the transition from error mitigation to error correction as we proceed from NISQ devices towards scalable quantum computing. 
This then complements presently active research in error suppression that combines the two techniques~\cite{suzuki2020quantum,lostaglio2021error,piveteau2021error,xiong2020sampling}. Beyond error suppression, quantum protocols in many diverse settings also share the structure of classical post-processing of quantum measurements --- from quantum metrology and illumination to hypothesis testing and stochastic analysis~\cite{lloyd2008enhanced,giovannetti2006quantum,audenaert2008asymptotic,binder2018practical,blank2021quantum}. Our framework --- suitably extended --- could thus identify new performance bounds in each of these settings.


\textit{Note added.}---During the completion of our manuscript, we became aware of an independent work by Wang \textit{et al.}~\cite{Wang2021can}, which showed a result related to our Theorem~\ref{thm:layer} on the exponential scaling of the maximum estimator spread.

\section*{Methods}

\textbf{Formal definition of $(Q,K)$-error mitigation} ---
Here, we give a formal definition of $(Q,K)$-error mitigation as a quantum operation.
Since POVM measurements in different experiments are independent of each other, the whole measurement process can be represented as a tensor product of each POVM. Then, the classical post-processing following the measurement is a classical-classical channel such that the expected value of the output will serve as an estimate of the desired expectation value. 
We can then formalize an error-mitigation process as a concatenation of these two maps. 

\begin{defn}[$(Q,K)$-error mitigation]\label{def:mitigation}
For an arbitrary observable $A$ satisfying $-\mbI/2\leq A \leq \mbI/2$, a $(Q,K)$-mitigation protocol --- involving $Q$ inputs and $K$ experiments --- is a concatenation of quantum-classical channel $\Lambda_A$ and classical-classical channel $\hat e_A$ as $\hat e_A\circ\Lambda_A$.
Here, $\Lambda_A$ has a form 
\bal
 \Lambda_A(\cdot)&=\sum_{\bf i}\Tr(\cdot\, M_{i^{(1)}}^{(1)}\otimes\dots\otimes M_{i^{(K)}}^{(K)}) \,\dm{\bf i}
 \label{eq:mitigation channel}
\eal
where $\{M_{i^{(k)}}^{(k)}\}$ is the POVM for the $k^{\mathrm{th}}$ experiment acting on $Q$ copies of $n$-qubit noisy states, and ${\bf i}\coloneqq i^{(1)}\dots i^{(K)}$ denotes a collection of measurement outcomes with $\ket{\bf i}=\ket{i^{(1)}\dots i^{(K)}}$ being a classical state acting on $K$ subsystems. The channel $\hat e_A$ implements a $K$-input classical function $e_A$ such that
\bal
 \sum_{\bf i}p_{\bf i} e_A({\bf i}) = \Tr(A\psi) + b_A(\psi)
\label{eq:estimate}
\eal
for some function $b_A(\psi)$ called bias, and 
\bal
 p_{\bf i}\coloneqq \prod_{k=1}^{K}\Tr[\mE_1^{(k)}(\psi)\otimes\cdots\otimes\mE_Q^{(k)}(\psi)\, M_{i^{(k)}}^{(k)}]
 \label{eq:outcome distribution}
\eal
is the probability of getting outcomes ${\bf i}=i^{(1)}\dots i^{(K)}$ for the input noisy states $\{\mE_q^{(k)}(\psi)\}_{q=1,k=1}^{Q,K}$. 
\end{defn}


\textbf{Proof of Theorem~\ref{thm:spread}} ---
The intuition behind Theorem~\ref{thm:spread} lies in the intimate relation between the effect of error mitigation and distinguishability of quantum states. 
Recall that the goal of quantum error mitigation is to estimate the expectation value of an arbitrary observable $A$ for an arbitrary ideal state $\psi$ only using the noisy state $\mE(\psi)$. 
Although $\Tr(A\mE(\psi))$ can deviate from $\Tr(A\psi)$, error mitigation correctly allows us to estimate $\Tr(A\psi)$, which appears to have eliminated noise effects. 
Since each error-mitigation strategy should also work for another state $\phi$, it should be able to remove the noise and estimate $\Tr(A\phi)$ out of $\Tr(A\mE(\phi))$.
Does this `removal' of noise imply that error mitigation can help distinguish $\mE(\psi)$ and $\mE(\phi)$?

\begin{figure}[h]
    \begin{tikzpicture}
\draw[semithick] (-3,0)--(3,0);
\draw[semithick] (-3,-2)--(3,-2);
\draw[thick,samples=100,domain=-0.5:1.5] plot(\x,{1.3*exp(-6*pow((\x-0.6),2)});
\draw[thick,samples=100,domain=-1.5:0.5] plot(\x,{1.3*exp(-6*pow((\x+0.6),2)});
\draw[thick,dashed](-0.6,1.5)--(-0.6,-1.2);
\draw[thick,dashed](0.6,1.5)--(0.6,-1.2);
\draw(-0.8,1.7) node {{\footnotesize $\mathrm{Tr}(A\mathcal E(\psi))$}};
\draw(0.8,1.7) node {{\footnotesize $\mathrm{Tr}(A\mathcal E(\phi))$}};
\draw[thick,samples=100,domain=-1:3] plot(\x,{-2+0.5*exp(-pow((\x-1),2)});
\draw[thick,samples=100,domain=-3:1] plot(\x,{-2+0.5*exp(-pow((\x+1),2)});
\draw[thick,dashed](-1,-0.7)--(-1,-2);
\draw[thick,dashed](1,-0.7)--(1,-2);

\draw(-1.4,-0.3) node {{\footnotesize $\mathrm{Tr}(A\psi)$}};
\draw(1.4,-0.3) node {{\footnotesize $\mathrm{Tr}(A\phi)$}};
\draw(-1.4,-2.3) node {{\footnotesize $\mathrm{Tr}(A\psi)$}};
\draw(1.4,-2.3) node {{\footnotesize $\mathrm{Tr}(A\phi)$}};
\draw[semithick](-1.3,0)--(-1.3,0.15);
\draw[semithick](1.3,0)--(1.3,0.15);
\draw[semithick](-1.3,-2)--(-1.3,-1.85);
\draw[semithick](1.3,-2)--(1.3,-1.85);
\draw[<-,double](-1,-1)--(-0.6,-1);
\draw[->,double](0.6,-1)--(1,-1);

\fill [black] plot [smooth,samples=100,domain=0:0.6] (\x,{1.3*exp(-6*pow((\x+0.6),2)}) -- (1,0) -- (0,0) -- (0,1);
\fill [black] plot [smooth,samples=100,domain=-0.6:0](\x,{1.3*exp(-6*pow((\x-0.6),2)})-- (0,0.2)--(0,0)--(-0.6,0);

\fill [cyan] plot [smooth,samples=100,domain=0:1] (\x,{-2+0.5*exp(-pow((\x+1),2)}) -- (1,-2) -- (0,-2) -- (0,-1.8);
\fill [cyan] plot [smooth,samples=100,domain=-1:0](\x,{-2+0.5*exp(-pow((\x-1),2)})-- (0,-1.8)--(0,-2)--(-1,-2);
\end{tikzpicture}
    \caption{{\bf Error mitigation and distinguishability.} The top schematic illustrates the probability distribution of an observable $A$ for two noisy states $\mE(\psi)$ and $\mE(\phi)$. The expectation values are shifted from the true values due to the noise effects. As in the bottom schematic, error mitigation converts them to other distributions whose expectation values are closer to the true values than the initial noisy distributions are. However, the converted distributions get broader, and the overlap between two distributions increases in general.}
    \label{fig:distinguish}
\end{figure}
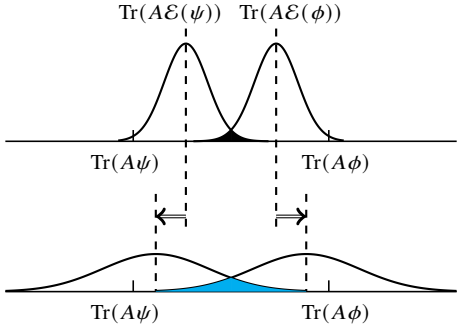

The subtlety of this question can be seen by looking at how quantum error mitigation works.
The estimation of $\Tr(A\mE(\psi))$ without error mitigation is carried out by making a measurement with respect to the eigenbasis of $A=\sum_a a\dm{a}$, which produces a probability distribution $p(a|\mE(\psi),A)$ over possible outcomes $\{a\}$.
Because of the noise, the expectation value of this distribution is shifted from $\Tr(A\psi)$.
Similarly, the same measurement for a state $\mE(\phi)$ produces a probability distribution $p(a|\mE(\phi),A)$, whose expectation value may also be shifted from $\Tr(A\phi)$. 
An error-mitigation protocol applies additional operations, measurements and classical post-processing to produce other probability distributions $p_{\rm EM}(a|\mE(\psi),A)$ and $p_{\rm EM}(a|\mE(\phi),A)$ whose expectation values get closer to the original ones. 
As a result, although the expectation values of the two error-mitigated distributions get separated from each other, they also get broader, which may increase the overlap between the two distributions, possibly making it even harder to distinguish two distributions. (See Fig.~\ref{fig:distinguish}.) 

One can see that this intuition that error mitigation does not increase the distinguishability is indeed right by looking at the whole error-mitigation process as a quantum channel. 
Then, the data-processing inequality implies that the distinguishability between any two states should not be increased by the application of quantum channels.    
This motivates us to rather use this observation as a basis to put a lower bound for the necessary overhead. 

Let us recall that the trace distance admits the following form
\bal
 D_{\rm tr}(\rho,\sigma)&= \frac{1}{2}\|\rho-\sigma\|_1\\
 &= \max_{0\leq M \leq \mbI}\Tr\left[M(\rho-\sigma)\right],
\label{eq:trace distance def app}
\eal
and similarly the local distinguishablity measure can be written as~\cite{matthews_distinguishability_2009}
\bal
 D_{\textrm{LM}}(\rho,\sigma)&= \max_{\{M_i\}\in \textrm{LM}}\frac{1}{2}\|\mM(\rho)-\mM(\sigma)\|_1\\
 &= \max_{\{M,\mbI-M\}\in \textrm{LM}_2 }\Tr[M(\rho-\sigma)]
\label{eq:LM def app}
\eal
where $\mathrm{LM}$ is the set of POVMs that take the form $M_{i^{(1)}}^{(1)}\otimes\dots\otimes M_{i^{(K)}}^{(K)}$, where $M_{i^{(k)}}^{(k)}$ represents some POVM local to system $S_k$, and $\textrm{LM}_2$ is the set of two-outcome measurements realized by local measurements together with classical post-processing. 
The second forms for the above measures particularly tell that they quantify how well two states can be distinguished by accessible quantum measurements. 
By definition, it is clear that 
\bal
 D_{\tr}(\rho,\sigma)\geq D_{\rm LM}(\rho,\sigma)
\label{eq:trace and LM inequality}
\eal
for all states $\rho$ and $\sigma$, and the inequality often becomes strict~\cite{Lami2017ultimate,Correa2021maximal}.

The local distinguishability measure satisfies the data-processing inequality under all local measurement channels. 
Namely, for all states $\rho$ and $\sigma$ defined on a composite system $\otimes_{k=1}^K S_k$, and for an arbitrary quantum-classical channel $\Lambda(\cdot)=\sum_i \Tr\left(\,\cdot\, M_{i^{(1)}}^{(1)}\otimes\cdots\otimes M_{i^{(K)}}^{(K)}\right)\dm{i^{(1)}\dots i^{(K)}}$,  
\bal
 D_{\textrm{LM}}(\Lambda(\rho),\Lambda(\sigma))&=\max_{\mM\in{\rm LM}}\frac{1}{2}\|\mM\circ\Lambda(\rho)-\mM\circ\Lambda(\sigma)\|_1\\
 &\leq \max_{\mM\in{\rm LM}}\frac{1}{2}\|\mM(\rho)-\mM(\sigma)\|_1\\
 &=D_{\textrm{LM}}(\rho,\sigma)
 \label{eq:data-processing LM}
\eal
where in the inequality we used that the set of local measurement channels is closed under concatenation.

Let us define 
\bal
\tilde\psi_Q^{(K)}&\coloneqq \otimes_{k=1}^{K}\otimes_{q=1}^{Q}\left[\mE_q^{(k)}(\psi)\right],\\
\tilde\phi_Q^{(K)}&\coloneqq \otimes_{k=1}^{K}\otimes_{q=1}^{Q}\left[\mE_q^{(k)}(\phi)\right].
\eal
Since the channel $\Lambda_A$ in Definition~\ref{def:mitigation} is a local measurement channel, we employ \eqref{eq:data-processing LM} to get
\bal
 D_{\textrm{LM}}\left(\tilde\psi_Q^{(K)},\tilde\phi_Q^{(K)}\right) &\geq D_{\textrm{LM}}\left(\Lambda_A\left(\tilde\psi_Q^{(K)}\right),\Lambda_A\left(\tilde\phi_Q^{(K)}\right)\right)\\
 &=D_{\textrm{LM}}(\hat p,\hat q)
 \label{eq:data processing}
\eal
where 
\bal
\hat p&=\sum_{\bf i}p_{\bf i}\,\dm{\bf i},\quad
\hat q&=\sum_{\bf i}q_{\bf i}\dm{\bf i}
\eal
and $p_{\bf i}$ and $q_{\bf i}$ are classical distributions defined in $\eqref{eq:outcome distribution}$ for $\psi$ and $\phi$ respectively, which satisfy
\bal
 \sum_{\bf i} p_{\bf i} e_A({\bf i}) &= \Tr(A\psi) + b_A(\psi),\\
 \sum_{\bf i} q_{\bf i} e_A(\bf i) &= \Tr(A\phi) + b_A(\phi).
 \label{eq:estimator for two states}
\eal

When $\hat p$ and $\hat q$ are tensor products of classical states, i.e., $\hat p=\hat p^{(1)}\otimes\dots\otimes \hat p^{(K)}$ and $\hat q=\hat q^{(1)}\otimes\dots\otimes \hat q^{(K)}$, it holds that 
\bal
D_{\textrm{LM}}(\hat p,\hat q)=D_{\rm tr}(\hat p,\hat q).
\label{eq:LM = trace}
\eal

This can be seen as follows. Let $M^\star$ be the optimal POVM element achieving the trace distance in \eqref{eq:trace distance def app}.
Then, we get 
\bal
D_{\rm tr}(\hat p,\hat q) &= \Tr[M^\star (\hat p-\hat q)]\\
&= \Tr[\Delta(M^\star) (\hat p-\hat q)]
\eal
where 
\bal
\Delta(\cdot)\coloneqq\sum_{\bf i} \dm{\bf i}\cdot \dm{\bf i}
\eal
is a classical dephasing channel.
The effective POVM element $\Delta(M^\star)$ has the form 
\bal
\Delta(M^\star)=\sum_{\bf i}\braket{{\bf i}|M^\star|{\bf i}}\dm{\bf i}.
\eal
Since each $\dm{\bf i}$ is a local POVM element and $0\leq \braket{{\bf i}|M^\star|{\bf i}}\leq 1$ because $0\leq M^\star\leq \mbI$, the two-outcome measurement $\{\Delta(M^\star),\mbI-\Delta(M^\star)\}$ can be realized by a local measurement and classical post-processing, and thus belongs to ${\rm LM_2}$. 
This, together with \eqref{eq:LM def app}, implies $D_{\tr}(\hat p,\hat q)\leq D_{\textrm{LM}}(\hat p,\hat q)$, and further combining \eqref{eq:trace and LM inequality} gives \eqref{eq:LM = trace}. 

Combining \eqref{eq:data processing} and \eqref{eq:LM = trace} gives  
\bal
 D_{\tr}(\hat p,\hat q)\leq D_{\rm LM}\left(\tilde\psi_Q^{(K)},\tilde\phi_Q^{(K)}\right).
\label{eq:trace LM data processing}
\eal

We now connect \eqref{eq:trace LM data processing} to the expression \eqref{eq:estimator for two states} of the expectation value and bias.  
Let us first suppose $\Tr(A\psi)+b_A(\psi)\geq \Tr(A\phi)+b_A(\phi)$. 
Let $\mI^\star\coloneqq\lset {\bf i} \sbar p_{\bf i}-q_{\bf i}\geq 0\rset$ and let $\bar \mI^\star$ be the complement set.
Let us also define $A'=A+\mbI/2$, which satisfies $0\leq A' \leq \mbI$ due to $-\mbI/2\leq A \leq \mbI/2$.
Then, we get
\begin{equation}\begin{aligned}
 &\Tr[A'(\psi-\phi)] + b_A(\psi)-b_A(\phi)\\
 &=\Tr[(A+\mbI/2)(\psi-\phi)] + b_A(\psi)-b_A(\phi)\\
 &=\Tr[A(\psi-\phi)] + b_A(\psi)-b_A(\phi)\\
 &=\sum_{\bf i} (p_{\bf i}-q_{\bf i}) e_A(\bf i)\\ 
 &\leq \sum_{\bf i\in\mI^\star}(p_{\bf i}-q_{\bf i})e_{A,\max} + \sum_{{\bf i}\in \bar\mI^\star} (p_{\bf i}-q_{\bf i})e_{A,\min}\\
 &=D_{\tr}(\hat p,\hat q)(e_{A,\max} - e_{A,\min})
\label{eq:bias and spectral spread}
\end{aligned}\end{equation}
where in the third line we used \eqref{eq:estimator for two states}, in the fourth line we used the maximum and minimum estimator values
\bal
e_{A,\max} \coloneqq \max_{\bf i}e_A({\bf i}),\quad e_{A,\min} \coloneqq \min_{\bf i}e_A({\bf i}),
\eal
and in the last line we used that 
\bal
\sum_{{\bf i}\in\bar\mI^\star}(p_{\bf i}-q_{\bf i})=- \sum_{{\bf i}\in\mI^\star}(p_{\bf i}-q_{\bf i})
\eal
and that the trace distance reduces to the total variation distance 
\bal
 D_{\tr}(\hat p,\hat q)=\sum_{i:p_i-q_i\geq 0}(p_i-q_i)
 \label{eq:trace distance classical def}
\eal
for all classical states $\hat p=\sum_i p_i \dm{i}$ and $\hat q=\sum_i q_i\dm{i}$.
Combining \eqref{eq:trace LM data processing} and \eqref{eq:bias and spectral spread}, we get 
\bal
 e_{A,\max}-e_{A,\min}\geq \frac{\Tr[A'(\psi-\phi)]+b_A(\psi)-b_A(\phi)}{D_{\rm LM}\left(\tilde\psi_Q^{(K)},\tilde\phi_Q^{(K)}\right)}.
\eal

On the other hand, if $\Tr(A\psi)+b_A(\psi)\leq \Tr(A\phi)+b_A(\phi)$, we flip the role of $\psi$ and $\phi$ to get 
\bal
 e_{A,\max}-e_{A,\min}\geq -\frac{\Tr[A'(\psi-\phi)]+b_A(\psi)-b_A(\phi)}{D_{\rm LM}\left(\tilde\psi_Q^{(K)},\tilde\phi_Q^{(K)}\right)}.
\eal
Defining $\Delta e_A\coloneqq e_{A,\max}-e_{A,\min}$, these two can be summarized as 
\bal
\Delta e_A \geq \frac{\left|\Tr[A'(\psi-\phi)]+b_A(\psi)-b_A(\phi)\right|}{D_{\rm LM}\left(\tilde\psi_Q^{(K)},\tilde\phi_Q^{(K)}\right)}.
\label{eq:spread fixed observable}
\eal

Optimizing over $A$, $\phi$, and $\psi$ on both sides, we reach
\bal
 \Delta e_{\max}&\geq  \max_{\substack{\psi,\phi\\-\mbI/2\leq A \leq \mbI/2}}\frac{|\Tr[A'(\psi-\phi)]+b_A(\psi)-b_A(\phi)|}{D_{\rm LM}\left(\tilde\psi_Q^{(K)},\tilde\phi_Q^{(K)}\right)}\\
 &=\max_{\substack{\psi,\phi\\-\mbI/2\leq A \leq \mbI/2}}\frac{\Tr[A'(\psi-\phi)]+b_A(\psi)-b_A(\phi)}{D_{\rm LM}\left(\tilde\psi_Q^{(K)},\tilde\phi_Q^{(K)}\right)}\\
&\geq\max_{\substack{\psi,\phi}}\frac{D_{\rm tr}(\psi,\phi)+b_{A^\star}(\psi)-b_{A^\star}(\phi)}{D_{\rm LM}\left(\tilde\psi_Q^{(K)},\tilde\phi_Q^{(K)}\right)}\\
&\geq\max_{\substack{\psi,\phi}}\frac{D_{\rm tr}(\psi,\phi)-2b_{\max}}{D_{\rm LM}\left(\tilde\psi_Q^{(K)},\tilde\phi_Q^{(K)}\right)}
\label{eq:spectrum bound}
\eal
where in the second line we used that we can always take the numerator positive by appropriately flipping $\psi$ and $\phi$, in the third line we fixed ${A'}^\star=A^\star+\mbI/2$ to the one that achieves the trace distance $\Tr[{A'}^\star(\psi-\phi)]=D_{\tr}(\psi,\phi)$ as in \eqref{eq:trace distance def app}, and in the fourth line we used the definition of $b_{\max}$.

\qed

\textbf{Measuring subfidelity} ---
To estimate the subfidelity \eqref{eq:subfidelity} for $n$-qubit states $\rho$ and $\sigma$, it suffices to measure the two quantities, $\Tr(\rho\sigma)$ and $\Tr(\rho\sigma\rho\sigma)$, which can be measured by a quantum computer~\cite{Ekert2002direct,Bacon2006efficient}. 
For readers' convenience, here we summarize several methods that can measure the subfidelity and see that the measurement can be done by a constant-depth quantum circuit. 

Let us begin by $\Tr(\rho\sigma)$.
Note that $\Tr(\rho\sigma)=\Tr(S\,\rho\otimes\sigma)$ where $S$ is the $n$-qubit SWAP operator defined by $S\ket{\psi}\otimes\ket{\phi}=\ket{\phi}\otimes\ket{\psi}$ with  $\ket{\psi}$ and $\ket{\phi}$ being arbitrary $n$-qubit pure states.
This can be famously measured by the SWAP test~\cite{Ekert2002direct} that uses one ancillary qubit and $n$-qubit SWAP gate controlled on the ancillary qubit. 
Since the $n$-qubit SWAP gate can be realized by swapping individual qubits, the SWAP test runs with $n$ uses of qubit SWAP gates controlled on the ancillary qubit, taking the circuit depth $n$.

One can significantly reduce the circuit depth by employing the destructive SWAP test~\cite{Garcia-Escartin2013swap}. 
Note that $\Tr(\rho\sigma)=\Tr(S_2^{\otimes n}\rho\otimes\sigma)$ where $S_2 \coloneqq \sum_{i,j=0}^1 \ketbra{ij}{ji}$ is the qubit SWAP operator. 
This is obtained by measuring $\rho\otimes\sigma$ with respect to the eigenbasis of $S_2^{\otimes n}$, which is just a tensor product of the eigenbasis of $S_2$.
Therefore, such a measurement can be accomplished by individually measuring a pair of qubits from $\rho$ and $\sigma$ with respect to the eigenbasis of $S_2$, for which one can use, e.g., Bell measurement.
These measurements can run in parallel and thus only needs a constant depth circuit with respect to $n$ (in fact, depth 2) that involves $n$ two-qubit gates. 

We remark that, at this point, we have already obtained a valid lower bound of $\Delta e_{\max}$ because the second term in \eqref{eq:subfidelity} is positive, only improving the lower bound. 
Nevertheless, evaluating the second term, which involves $\Tr(\rho\sigma\rho\sigma)$, can significantly improve the bound particularly when $\rho$ and $\sigma$ are highly noisy and their purity is small. 

$\Tr(\rho\sigma\rho\sigma)$ can be measured by a similar strategy to the one for $\Tr(\rho\sigma)$ with two copies of $\rho$ and $\sigma$. 
Instead of the SWAP operator $S$, consider the CYCLE operator $C$ defined as $C\left(\otimes_{i=1}^4\ket{\psi_i}\right)=\otimes_{i=1}^4\ket{\psi_{i+1}}$ where $\ket{\psi_i}, i=1,2,3,4$ is an arbitrary $n$-qubit pure state with $\ket{\psi_5}\coloneqq \ket{\psi_1}$.
Then, it is straightforward to check that $\Tr(\rho\sigma\rho\sigma)=\Tr(C\,\rho\otimes\sigma\otimes\rho\otimes\sigma)$. 
This can be measured by a generalization of the SWAP test where CYCLE gate $C$ is controlled on the single ancillary qubit. 
Similarly to the case of SWAP, the CYCLE gate $C$ can be decomposed into $C=C_2^{\otimes n}$ where $k^{\rm th}$ $C_2$ gate (for any $k=1,\dots, n$) acts on the four-qubit state that consists of the $k^{\rm th}$ qubit of $\rho$, $\sigma$, $\rho$, and $\sigma$.
Since $C_2$ can be realized by three SWAP gates, one can measure $\Tr(\rho\sigma\rho\sigma)$ with $3n$ uses of qubit-SWAP gates controlled on the ancillary qubit, taking the circuit depth $3n$.

Similarly to the case of $\Tr(\rho\sigma)$, we can realize a significant reduction in the circuit depth by making the measurement destructive. 
All we have to do is to measure individual four-qubit states that each $C_2$ gate acts on with respect to the eigenbasis of $C_2$.
Since the measurement of each $C_2$ can be run in parallel and each measurement circuit has a depth independent of $n$, this results in a constant-depth circuit that measures $C=C_2^{\otimes n}$.

We note the apparent similarity between the construction above and the circuit used in virtual distillation~\cite{koczor2021exponential,huggins2021virtual}. 
In particular, the strategy of destructive measurement was extensively discussed in Ref.~\cite{huggins2021virtual}.
It is interesting to see that a construction that is highly relevant to a specific error-mitigation protocol provides a bound applicable to a general class of error-mitigation protocols. 


\textbf{Applications to other error-mitigation protocols} ---
Here, we discuss how our framework can be applied to other two prominent error-mitigation protocols, noise extrapolation and virtual distillation.

Extrapolation methods~\cite{Temme2017error,Li2017efficient} are used in scenarios where there is no clear analytical noise model. These strategies consider a family of noise channels $\{\mN_\xi\}_\xi$, where $\xi$ corresponds to the noise strength.
The assumption here is that the description of $\mN_\xi$ is unknown, but we have the ability to `boost' $\xi$ such that $\xi\geq \tilde\xi$ where $\tilde\xi$ is the noise strength present in some given noisy circuit. The idea is that by studying how the expectation value of an observable depends on $\xi$, we can extrapolate what its value would be if $\xi = 0$. 
In particular, the $R^{\mathrm{th}}$ order Richardson extrapolation method work as follows. 
Let us take constants $\{\gamma_r\}_{r=0}^R$ and $\{c_r\}_{r=0}^R$ with $1=c_0<c_1<\dots<c_R\leq 1/\tilde\xi$ such that  

\bal
 \sum_{r=0}^R \gamma_r = 1,\ \ \sum_{r=0}^R \gamma_r c_r^t = 0\quad t=1,\dots,R.
 \label{eq:extrapolation condition}
\eal
Using these constants, one can show that 
\bal
 \sum_{r=0}^{R} \gamma_r \Tr[A\mN_{c_r\tilde\xi}(\psi)]=\Tr(A\psi)+b_A(\psi)
 \label{eq:extrapolation richardson}
\eal
where $b_A(\psi)=\mO(\tilde\xi^{R+1})$.
This allows us to estimate the true expectation value using noisy states under multiple noise levels, as long as $\tilde\xi$ is sufficiently small. 

Richardson extrapolation is an instance of $(1,R+1)$-error mitigation. In particular, we have 
\bal
\mE^{(k)}=\mN_{c_{k-1}\tilde\xi}\quad k=1,\dots,R+1
\eal
in Definition~\ref{def:mitigation}.
For an observable $A=\sum_a a\Pi_{a}$ where $\Pi_a$ is the projector corresponding to measuring outcome $a$, the POVMs $\{M_{a^{(k)}}^{(k)}\}_{k=1}^{R+1}$ and classical estimator function $e_A$ take the forms 
\bal
M_{a^{(k)}}^{(k)}=\Pi_{a^{(k)}}\quad k=1,\dots,R+1,
\eal
\bal
 e_A(a^{(1)},\dots,a^{(R+1)}) = \sum_{k=1}^{R+1} \gamma_{k-1} a^{(k)},
\eal
where $\{\gamma_k\}_{k=0}^{R}$ are the constants determined by \eqref{eq:extrapolation condition}.
One can easily check that plugging the above expressions in the form of Definition~\ref{def:mitigation} leads to \eqref{eq:extrapolation richardson}. 

Because of the constraint $-\mbI/2\leq A \leq \mbI/2$, every eigenvalue $a$ satisfies $-1/2\leq a \leq 1/2$. 
This implies that
\bal
 e_{A,\max}&\leq \frac{1}{2}\sum_{r:\gamma_r\geq 0}\gamma_{r}-\frac{1}{2}\sum_{r:\gamma_r< 0}\gamma_{r}\\
 &=\frac{1}{2}\sum_{r=0}^R|\gamma_{r}|
\eal
and 
\bal
 e_{A,\min}&\geq -\frac{1}{2}\sum_{r:\gamma_r\geq 0}\gamma_{r}+\frac{1}{2}\sum_{r:\gamma_r< 0}\gamma_{r}\\
 &=-\frac{1}{2}\sum_{r=0}^R|\gamma_{r}|,
\eal
leading to $\Delta e_{\max}\leq \sum_{r=0}^R |\gamma_r|$. 
On the other hand, any observable $A$ having $\pm 1/2$ eigenvalues saturates this inequality. 
Therefore, we get the exact expression of the maximum spread for the extrapolation method as
\bal
 \Delta e_{\max}^{\rm EX}=\sum_{r=0}^R |\gamma_r|.
 \label{eq:extrapolation spread formula}
\eal

Next, we discuss virtual distillation~\cite{koczor2021exponential,huggins2021virtual}, which is an example of $(Q,1)$-error mitigation.
Let $\psi$ be an ideal pure output state from a quantum circuit. 
We consider a scenario where the noise in the circuit acts as an effective noise channel $\mE$ that brings the ideal state to a noisy state of the form
\bal
 \mE(\psi)=\lambda \psi + \sum_{k=2}^{d} \lambda_k \psi_k
 \label{eq:virtual distillaton output}
\eal
for a certain $\{\lambda_k\}_{k=1}^d$, where $d$ is the dimension of the system and $\{\psi_k\}_{k=1}^{d}$ constructs an orthonormal basis with $\psi_1\coloneqq \psi$.
We also assume that $\lambda$ is given as pre-knowledge.
This form reflects the intuition that, as long as the noise is sufficiently small, the dominant eigenvector should be close to the ideal state $\psi$.
For a more detailed analysis of the form of this spectrum, we refer readers to Ref.~\cite{koczor2021dominant}.

The $Q$-copy virtual distillation algorithm aims to estimate $\Tr(W\psi)$ for a unitary observable $W$ satisfying $W^2=\mbI$ (e.g., Pauli operators) by using $Q$ copies of $\mE(\psi)$.
The mitigation circuit consists of a controlled permutation and unitary $W$, followed by a measurement on the control qubit with the Hadamard basis.   
The probability of getting outcome 0 (projecting onto $\dm{+}$) is 
\bal
 p_0 &= \frac{1}{2}\left(1+\Tr\left[W\mE(\psi)^Q\right]\right)\\
 &=\frac{1}{2}\left[1+\lambda^Q\Tr(W\psi)+\sum_{k=2}^d\lambda_k^Q\Tr(W\psi_k)\right].
 \label{eq:virtual prob}
\eal
This implies that 
\bal
(2p_0-1)\lambda^{-Q} = \Tr(W\psi)+\sum_{k=2}^d \left(\frac{\lambda_k}{\lambda}\right)^Q\Tr(W\psi_k),
\label{eq:virtual bias}
\eal
providing a way of estimating $\Tr(W\psi)$ with the bias $|\sum_{k=2}^d(\lambda_k/\lambda)^Q\Tr(W\psi_k)|\leq \sum_{k=2}^d (\lambda_k/\lambda)^Q$.

We can see that this protocol fits into our framework with $K=1$ and $\mE_q=\mE$ for $q=1,\dots,Q$ as follows. 
For an arbitrary observable $A$, we can always find a decomposition with respect to the Pauli operators $\{P_i\}$ as 
\bal
 A=\sum_i c_i P_i
\eal
for some set of real numbers $\{c_i\}$.
We now apply the virtual distillation circuit for $P_i$ at probability $|c_i|/\sum_j|c_j|$ and --- similarly to the case of probabilistic error cancellation --- employ an estimator function defined as 
\bal
 e_A(i0)&\coloneqq \gamma \mathrm{sgn}(c_i) \lambda^{-Q}\\
 e_A(i1)&\coloneqq -\gamma \mathrm{sgn}(c_i) \lambda^{-Q}
 \label{eq:virtual estimator}
\eal
with $\gamma\coloneqq\sum_i|c_i|$, where we treat $i$ as a part of the measurement outcome. 
Then, we get 
\bal
 \sum_i \left[p_{i0}\, e_A(i0) +  p_{i1} e_A(i1)\right]=\Tr(A\psi) + b_A(\psi)
\eal
where $p_{i0}$ is the probability \eqref{eq:virtual prob} with $W=P_i$ multiplied by $|c_i|/\sum_j|c_j|$, $p_{i1}=1-p_{i0}$, and $b_A(\psi)\coloneqq\sum_{k=2}^d (\lambda_k/\lambda)^Q\Tr(A\psi_k)$.
Optimizing over observables $-\mbI/2\leq A\leq \mbI/2$, we have 
\bal
 \Delta e_{\max}^{\rm VD}=\max\lset 2\lambda^{-Q}\sum_i|c_i|\sbar -\mbI/2\leq \sum_i c_i P_i \leq \mbI/2\rset
 \label{eq:virtual performance spread}
\eal
and 
\bal
b_{\max}^{\rm VD} = \sum_{k=2}^d \frac{1}{2}\left(\frac{\lambda_k}{\lambda}\right)^Q.
\label{eq:virtual performance bias}
\eal


\section*{Data Availability}

No datasets were generated or analyzed during the current study.

\section*{Code availability} 

Source codes used to generate the plots are available from the corresponding author upon request.


\bibliographystyle{apsrmp4-2}
\bibliography{myref}


\begin{acknowledgments}

We thank Yuichiro Matsuzaki, Yuuki Tokunaga, Hideaki Hakoshima, Kaoru Yamamoto, Jayne Thompson, and Francesco Buscemi for fruitful discussions, and Kento Tsubouchi for pointing out an error in a preliminary version of the manuscript. This work is supported by the Singapore Ministry of Education Tier 1  Grant RG162/19 and RG146/20, the National Research Foundation under its Quantum Engineering Program NRF2021-QEP2-02-P06, the Singapore Ministry of Education Tier 2 Project MOE-T2EP50221-0005 and the FQXi-RFP-IPW-1903 project, 'Are quantum agents more energetically efficient at making predictions?' from the Foundational Questions Institute, Fetzer Franklin Fund, a donor advised fund of Silicon Valley Community Foundation, and the Lee Kuan Yew Postdoctoral Fellowship at Nanyang Technological University Singapore. Any opinions, findings and conclusions or recommendations expressed in this material are those of the author(s) and do not reflect the views of National Research Foundation or the Ministry of Education, Singapore. S.E. is supported by Moonshot R\&D, JST, Grant No.\,JPMJMS2061; MEXT Q-LEAP Grant No.\,JPMXS0120319794, and PRESTO, JST, Grant No.\, JPMJPR2114. S.M. would like to take this opportunity to thank the “Nagoya University Interdisciplinary Frontier Fellowship” supported by JST and Nagoya University.
\end{acknowledgments}


\section*{Competing Interests}

The authors declare no competing interests. 

\section*{Author Contribution}

R.T., S.E., and S.M. came up with a preliminary idea on connecting the state distinguishability to error mitigation. R.T. conceived the project, obtained the main results, and wrote the manuscript draft. S.E. proposed a way of directly estimating a lower bound of the fidelity bound on a quantum computer, which eventually resulted in the subfidelity bound. S.M. contributed to the analysis to compare the subfidelity bound to the trace-distance bound. R.T. and M.G. wrote the manuscript. All authors contributed to the interpretation and discussion of the results.




\clearpage
\newgeometry{hmargin=1.2in,vmargin=0.8in}

\setcounter{section}{0}
\setcounter{equation}{0}
\setcounter{figure}{0}
\renewcommand{\thesection}{\arabic{section}}
\renewcommand{\theequation}{S\arabic{equation}}
\renewcommand{\figurename}{Supplementary Figure}

\titleformat{\section}
  {\centering\normalfont\bfseries\large}
  {Supplementary Note \thesection:}{.5em}{}

\widetext


\vspace{-.5cm}


\section{Proof of Theorem~\ref{thm:layer}}\label{app:layer}

\begin{proof}

For an arbitrary unitary channel $\mV$, Eq.~\eqref{eq:spread bound} in Theorem~\ref{thm:spread} in the main text can also be written as

\begin{equation}\begin{aligned}
 \Delta e_{\max}&\geq\max_{\psi,\phi}\frac{D_{\tr}(\psi,\phi)-2b_{\max}}{D_{\rm LM}\left(\otimes_{k=1}^{K}\otimes_{q=1}^{Q}\mE_q^{(k)}(\psi),\otimes_{k=1}^{K}\otimes_{q=1}^{Q}\mE_q^{(k)}(\phi)\right)}\\
 &=\max_{\psi,\phi}\frac{D_{\tr}(\mV^\dagger(\psi),\mV^\dagger(\phi))-2b_{\max}}{D_{\rm LM}\left(\otimes_{k=1}^{K}\otimes_{q=1}^{Q}\mE_q^{(k)}\circ\mV(\mV^\dagger(\psi)),\otimes_{k=1}^{K}\otimes_{q=1}^{Q}\mE_q^{(k)}\circ\mV(\mV^\dagger(\phi))\right)}\\
 &=\max_{\psi_{\rm in},\phi_{\rm in}}\frac{D_{\tr}(\psi_{\rm in},\phi_{\rm in})-2b_{\max}}{D_{\rm LM}\left(\otimes_{k=1}^{K}\otimes_{q=1}^{Q}\mE_q^{(k)}\circ\mV(\psi_{\rm in}),\otimes_{k=1}^{K}\otimes_{q=1}^{Q}\mE_q^{(k)}\circ\mV(\phi_{\rm in})\right)},
 \label{eq:spread inequality with unitary}
\end{aligned}\end{equation}
where in the second line we used the unitary invariance of the trace distance, and in the third line we changed the variables as $\mV^\dagger(\psi)\to\psi_{\rm in}$, $\mV^\dagger(\phi)\to\phi_{\rm in}$ and used the fact that the application of a fixed unitary does not affect the optimization taken over all states.

The noise model for layered circuits typically assumes the application of a noise channel after each layer. 
Let $\mN_{q,l}^{(k)}$ be a noise channel after the $l^{\mathrm{th}}$ layer for the $q^{\mathrm{th}}$ input in the $k^{\mathrm{th}}$ experiment. 
Then, a noisy circuit for the $q^{\mathrm{th}}$ input in the $k^{\mathrm{th}}$ experiment is described by 
\bal
\mN_{q, L}^{(k)}\circ\mU_{L}\circ\cdots\circ\mN_{q, 1}^{(k)}\circ\mU_1
\eal
while the ideal output given input state $\psi_{\rm in}$ is $\psi=\mU_{L}\circ\cdots\circ\mU_1(\psi_{\rm in})$. 
The effective noise channel for the $q^{\mathrm{th}}$ input in the $k^{\mathrm{th}}$ experiment is then given by 
\bal
\mE_{q}^{(k)}=\mN_{q,L}^{(k)}\circ\mU_{L}\circ\cdots\circ\mN_{q,1}^{(k)}\circ\mU_1\circ\mU_1^\dagger\circ\mU_2^\dagger\circ\cdots\circ\mU_{L}^\dagger.
\label{eq:effective noise layer}
\eal

Plugging \eqref{eq:effective noise layer} into \eqref{eq:spread inequality with unitary} while taking $\mV=\mU_L\circ\cdots\circ\mU_1$, we get

\begin{equation}\begin{aligned}
 \Delta e_{\max}&\geq\max_{\psi_{\rm in},\phi_{\rm in}}\frac{D_{\tr}(\psi_{\rm in},\phi_{\rm in})-2b_{\max}}{D_{\rm LM}\left(\otimes_{k=1}^{K}\otimes_{q=1}^{Q}\prod_{l=1}^{L}\left[\mN_{q,l}^{(k)}\circ\mU_{l}\right](\psi_{\rm in}),\otimes_{k=1}^{K}\otimes_{q=1}^{Q}\prod_{l=1}^{L}\left[\mN_{q,l}^{(k)}\circ\mU_{l}\right](\phi_{\rm in})\right)},
 \label{eq:multiple}
\end{aligned}\end{equation}
where we used the notation  
\bal
 \prod_{l=1}^{L}\left[\mN_{q,l}^{(k)}\circ\mU_{l}\right]\coloneqq\mN_{q,L}^{(k)}\circ\mU_{L}\circ\cdots\circ\mN_{q,1}^{(k)}\circ\mU_{1}
\eal

The denominator of the right-hand side of \eqref{eq:multiple} can be bounded as
\bal
&D_{\rm LM}\left(\otimes_{k=1}^{K}\otimes_{q=1}^{Q}\prod_{l=1}^{L}\left[\mN_{q,l}^{(k)}\circ\mU_{l}\right](\psi_{\rm in}),\otimes_{k=1}^{K}\otimes_{q=1}^{Q}\prod_{l=1}^{L}\left[\mN_{q,l}^{(k)}\circ\mU_{l}\right](\phi_{\rm in})\right)\\ &\leq D_{\rm tr}\left(\otimes_{k=1}^{K}\otimes_{q=1}^{Q}\prod_{l=1}^{L}\left[\mN_{q,l}^{(k)}\circ\mU_{l}\right](\psi_{\rm in}),\otimes_{k=1}^{K}\otimes_{q=1}^{Q}\prod_{l=1}^{L}\left[\mN_{q,l}^{(k)}\circ\mU_{l}\right](\phi_{\rm in})\right)\\
&\leq  D_{\rm tr}\left(\otimes_{k=1}^{K}\otimes_{q=1}^{Q}\prod_{l=1}^{L}\left[\mN_{q,l}^{(k)}\circ\mU_{l}\right](\psi_{\rm in}),\frac{\mbI}{2^{KQn}}\right) + D_{\rm tr}\left(\otimes_{k=1}^{K}\otimes_{q=1}^{Q}\prod_{l=1}^{L}\left[\mN_{q,l}^{(k)}\circ\mU_{l}\right](\phi_{\rm in}),\frac{\mbI}{2^{KQn}}\right)\\
&\leq  \sum_{k=1}^{K}\left[D_{\rm tr}\left(\otimes_{q=1}^{Q}\prod_{l=1}^{L}\left[\mN_{q,l}^{(k)}\circ\mU_{l}\right](\psi_{\rm in}),\frac{\mbI}{2^{Qn}}\right) + D_{\rm tr}\left(\otimes_{q=1}^{Q}\prod_{l=1}^{L}\left[\mN_{q,l}^{(k)}\circ\mU_{l}\right](\phi_{\rm in}),\frac{\mbI}{2^{Qn}}\right)\right] 
\label{eq:layer denominator bound}
\eal
where the first inequality is due to \eqref{eq:trace and LM inequality} in Methods, the second inequality is due to the triangle inequality, and in the last line we bounded each term by sequentially applying the triangle inequality as
\begin{equation}\begin{aligned}
&D_{\rm tr}\left(\otimes_{k=1}^{K}\otimes_{q=1}^{Q}\prod_{l=1}^{L}\left[\mN_{q,l}^{(k)}\circ\mU_{l}\right](\psi_{\rm in}),\frac{\mbI}{2^{KQn}}\right)\\
&\leq D_{\rm tr}\left(\otimes_{k=1}^{K}\otimes_{q=1}^{Q}\prod_{l=1}^{L}\left[\mN_{q,l}^{(k)}\circ\mU_{l}\right](\psi_{\rm in}),\frac{\mbI}{2^{Qn}}\otimes_{k=2}^{K}\otimes_{q=1}^{Q}\prod_{l=1}^{L}\left[\mN_{q,l}^{(k)}\circ\mU_{l}\right](\psi_{\rm in})\right)\\
&\quad+ D_{\rm tr}\left(\frac{\mbI}{2^{Qn}}\otimes_{k=2}^{K}\otimes_{q=1}^{Q}\prod_{l=1}^{L}\left[\mN_{q,l}^{(k)}\circ\mU_{l}\right](\psi_{\rm in}),\frac{\mbI}{2^{Qn}}\otimes\frac{\mbI}{2^{Qn(K-1)}}\right)\\
&=D_{\rm tr}\left(\otimes_{q=1}^{Q}\prod_{l=1}^{L}\left[\mN_{q,l}^{(1)}\circ\mU_{l}\right](\psi_{\rm in}),\frac{\mbI}{2^{Qn}}\right)+ D_{\rm tr}\left(\otimes_{k=2}^{K}\otimes_{q=1}^{Q}\prod_{l=1}^{L}\left[\mN_{q,l}^{(k)}\circ\mU_{l}\right](\psi_{\rm in}),\frac{\mbI}{2^{Qn(K-1)}}\right)\\
&\leq \dots\\
&\leq \sum_{k=1}^{K}D_{\rm tr}\left(\otimes_{q=1}^{Q}\prod_{l=1}^{L}\left[\mN_{q,l}^{(k)}\circ\mU_{l}\right](\psi_{\rm in}),\frac{\mbI}{2^{Qn}}\right), 
\end{aligned}\end{equation}
and similarly for the second term.
The last expression in \eqref{eq:layer denominator bound} can be further upper bounded as
\begin{equation}\begin{aligned}
 &\leq \sqrt{\frac{\ln 2}{2}} \sum_{k=1}^K\left(\sqrt{ S\left(\otimes_{q=1}^{Q}\prod_{l=1}^{L}\left[\mN_{q,l}^{(k)}\circ\mU_{l}\right](\psi_{\rm in})\,\Big\|\,\frac{\mbI}{2^{Qn}}\right)}+\sqrt{ S\left(\otimes_{q=1}^{Q}\prod_{l=1}^{L}\left[\mN_{q,l}^{(k)}\circ\mU_{l}\right](\phi_{\rm in})\,\Big\|\,\frac{\mbI}{2^{Qn}}\right)}\right),
 \label{eq:bound after Pinsker}
\end{aligned}\end{equation}
where we used the quantum Pinsker's inequality~\cite{Hiai1981sufficiency} 
\bal
 D_{\tr}(\rho,\sigma)\leq \sqrt{\frac{\ln 2}{2}}\,\sqrt{S(\rho\|\sigma)}
\eal
for all states $\rho$, $\sigma$, where $S(\rho\|\sigma)\coloneqq \Tr(\rho\log \rho)-\Tr(\rho\log\sigma)$ is the relative entropy.

We now recall the result in Ref.~\cite{Kastoryano2013quantum} (see also \cite{MullerHermes2016relative}), which evaluates the entropy increase due to the local depolarizing noise. 

\begin{lem}[\cite{Kastoryano2013quantum}]\label{lem:relative entropy decay}
Let $\mD_\epsilon(\rho)=(1-\epsilon)\rho+\epsilon\mbI/2$ be a qubit depolarizing channel. Then, for an arbitrary $n$-qubit state $\rho_n$, it holds that 
\bal
 S\left(\left({\mD_\epsilon}\right)^{\otimes n}(\rho_n)\,\Big\|\,\mbI/d^n\right)\leq (1-\epsilon)^2 S(\rho_n\,\|\,\mbI/d^n). 
\eal
\end{lem}

Then, for $\mN_{q,l}^{(k)}=\mD_{\epsilon_k}^{\otimes n}$ (see also Supplementary Figure~\ref{fig:noisy_layered_multicopies}), we get
\bal
 S\left(\otimes_{q=1}^{Q}\prod_{l=1}^{L}\left[\mN_{q,l}^{(k)}\circ\mU_{l}\right](\psi_{\rm in})\,\Big\|\,\frac{\mbI}{2^{Qn}}\right)&=S\left(\prod_{l=1}^{L}\left[\mD_{\epsilon_k}^{\otimes Qn}\circ\mU_{l}^{\otimes Q}\right](\psi_{\rm in}^{\otimes Q})\,\Big\|\,\frac{\mbI}{2^{Qn}}\right)\\
 &\leq (1-\epsilon_k)^2\,S\left(\mU_L^{\otimes Q}\prod_{l=2}^{L}\left[\mD_{\epsilon_k}^{\otimes Qn}\circ\mU_{l}^{\otimes Q}\right](\psi_{\rm in}^{\otimes Q})\,\Big\|\,\frac{\mbI}{2^{Qn}}\right)\\
 &=(1-\epsilon_k)^2\,S\left(\prod_{l=2}^{L}\left[\mD_{\epsilon_k}^{\otimes Qn}\circ\mU_{l}^{\otimes Q}\right](\psi_{\rm in}^{\otimes Q})\,\Big\|\,\frac{\mbI}{2^{Qn}}\right)\\
 &\leq (1-\epsilon_k)^{2L}\,S\left(\psi_{\rm in}^{\otimes Q}\,\Big\|\,\frac{\mbI}{2^{Qn}}\right)\\
 &\leq (1-\epsilon_k)^{2L} Qn,
 \label{eq:bound on relative entropy}
\eal
where the second line follows from Lemma~\ref{lem:relative entropy decay}, the third line is due to the unitary invariance of the relative entropy, in the fourth line we sequentially applied the same argument for $L$ times, and the fifth line is from the upper bound of the relative entropy, which is saturated by pure state $\psi_{\rm in}$. 

\begin{figure}
    \centering
    \begin{tikzpicture}
    \draw(-0.5,1.2)node{$\psi_{\rm in}$};
    
    \draw(0,0.3)--(0.5,0.3);
    \draw(0,1.4)--(0.5,1.4);
    \draw(0,2.1)--(0.5,2.1);
    \draw[thick,loosely dotted](0.25,0.5)--(0.25,1.2);
    \draw(0.5,0)rectangle(1.3,2.4);
    \draw(0.9,1.2)node{$\mathcal U_1$};
    \draw(1.3,0.3)--(1.8,0.3);
    \draw(1.3,1.4)--(1.8,1.4);
    \draw(1.3,2.1)--(1.8,2.1);
    \draw(1.8,0)rectangle(2.4,0.6);
    \draw(1.8,1.1)rectangle(2.4,1.7);
    \draw(1.8,1.8)rectangle(2.4,2.4);
    \draw(2.1,0.3)node{$\mathcal D_{\epsilon_k}$};
    \draw(2.1,1.4)node{$\mathcal D_{\epsilon_k}$};
    \draw(2.1,2.1)node{$\mathcal D_{\epsilon_k}$};
    \draw[thick,dotted](2.1,0.7)--(2.1,1.0);
    \draw(2.4,0.3)--(2.9,0.3);
    \draw(2.4,1.4)--(2.9,1.4);
    \draw(2.4,2.1)--(2.9,2.1);
    
    \draw[thick,loosely dotted](3,1.2)--(3.9,1.2);
    
    \draw(4,0.3)--(4.5,0.3);
    \draw(4,1.4)--(4.5,1.4);
    \draw(4,2.1)--(4.5,2.1);
    \draw(4.5,0)rectangle(5.3,2.4);
    \draw(4.9,1.2)node{$\mathcal U_L$};
    \draw(5.3,0.3)--(5.8,0.3);
    \draw(5.3,1.4)--(5.8,1.4);
    \draw(5.3,2.1)--(5.8,2.1);
    \draw(5.8,0)rectangle(6.4,0.6);
    \draw(5.8,1.1)rectangle(6.4,1.7);
    \draw(5.8,1.8)rectangle(6.4,2.4);
    \draw(6.1,0.3)node{$\mathcal D_{\epsilon_k}$};
    \draw(6.1,1.4)node{$\mathcal D_{\epsilon_k}$};
    \draw(6.1,2.1)node{$\mathcal D_{\epsilon_k}$};
    \draw[thick,dotted](6.1,0.7)--(6.1,1.0);
    \draw(6.4,0.3)--(6.9,0.3);
    \draw(6.4,1.4)--(6.9,1.4);
    \draw(6.4,2.1)--(6.9,2.1);
    
    \draw(8,1.2)node{$q=Q$};
    
    \draw[thick,loosely dotted](1.45,2.6)--(1.45,3.5);
    \draw[thick,loosely dotted](3.5,2.6)--(3.5,3.5);
    \draw[thick,loosely dotted](5.45,2.6)--(5.45,3.5);
    
    \draw(0.9,-0.5)node{$\underbrace{\quad\qquad}_{\mathcal U_1^{\otimes Q}}$};
    \draw(2.1,-0.5)node{$\underbrace{\qquad}_{\mathcal D_{\epsilon_k}^{\otimes Qn}}$};
    \draw(4.9,-0.5)node{$\underbrace{\quad\qquad}_{\mathcal U_L^{\otimes Q}}$};
    \draw(6.1,-0.5)node{$\underbrace{\qquad}_{\mathcal D_{\epsilon_k}^{\otimes Qn}}$};
    
    \draw[thick,loosely dotted](8,2.6)--(8,3.5);
    
    \draw(-0.5,4.9)node{$\psi_{\rm in}$};
    
    \draw(0,4)--(0.5,4);
    \draw(0,5.1)--(0.5,5.1);
    \draw(0,5.8)--(0.5,5.8);
    \draw(0.5,3.7)rectangle(1.3,6.1);
    \draw(0.9,4.9)node{$\mathcal U_1$};
    \draw(1.3,4)--(1.8,4);
    \draw(1.3,5.1)--(1.8,5.1);
    \draw(1.3,5.8)--(1.8,5.8);
    \draw(1.8,3.7)rectangle(2.4,4.3);
    \draw(1.8,4.8)rectangle(2.4,5.4);
    \draw(1.8,5.5)rectangle(2.4,6.1);
    \draw(2.1,4)node{$\mathcal D_{\epsilon_k}$};
    \draw(2.1,5.1)node{$\mathcal D_{\epsilon_k}$};
    \draw(2.1,5.8)node{$\mathcal D_{\epsilon_k}$};
    \draw(2.4,4)--(2.9,4);
    \draw(2.4,5.1)--(2.9,5.1);
    \draw(2.4,5.8)--(2.9,5.8);
    
    \draw[thick,loosely dotted](3,4.9)--(3.9,4.9);
    
    \draw(4,4)--(4.5,4);
    \draw(4,5.1)--(4.5,5.1);
    \draw(4,5.8)--(4.5,5.8);
    \draw(4.5,3.7)rectangle(5.3,6.1);
    \draw(4.9,4.9)node{$\mathcal U_L$};
    \draw(5.3,4)--(5.8,4);
    \draw(5.3,5.1)--(5.8,5.1);
    \draw(5.3,5.8)--(5.8,5.8);
    \draw(5.8,3.7)rectangle(6.4,4.3);
    \draw(5.8,4.8)rectangle(6.4,5.4);
    \draw(5.8,5.5)rectangle(6.4,6.1);
    \draw(6.1,4)node{$\mathcal D_{\epsilon_k}$};
    \draw(6.1,5.1)node{$\mathcal D_{\epsilon_k}$};
    \draw(6.1,5.8)node{$\mathcal D_{\epsilon_k}$};
    \draw(6.4,4)--(6.9,4);
    \draw(6.4,5.1)--(6.9,5.1);
    \draw(6.4,5.8)--(6.9,5.8);
    
    \draw(8,4.9)node{$q=1$};

    \end{tikzpicture}
    \caption{Noisy layered circuit for the $k^{\mathrm{th}}$ experiment under local depolarizing noise $\mD_{\epsilon_k}$. Each experiment contains $Q$ copies of the noisy circuit.}
    \label{fig:noisy_layered_multicopies}
\end{figure}
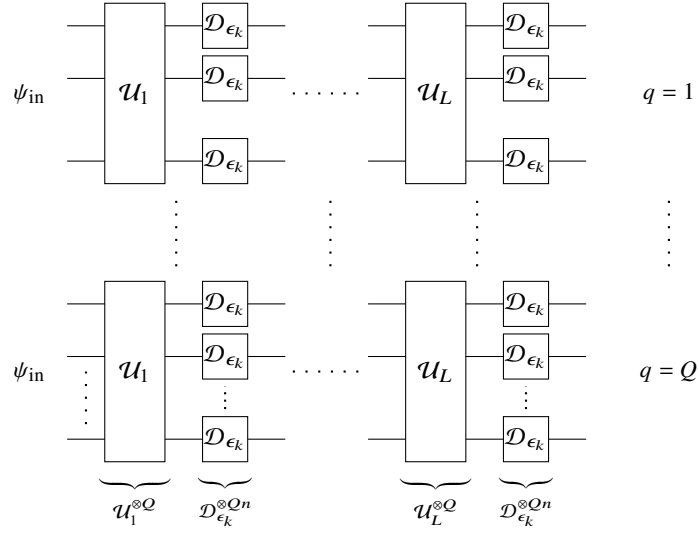

Using \eqref{eq:bound on relative entropy}, we can put a further bound on \eqref{eq:bound after Pinsker} as  

\begin{equation}\begin{aligned}
 &\leq \sqrt{2\ln 2} \sqrt{Qn} \sum_{k=1}^K (1-\epsilon_k)^{L}\\
 &\leq \sqrt{2\ln 2} \sqrt{Qn} K\, (1-\epsilon_{\min})^{L},
 \label{eq:upper bound of trace distance exponential}
\end{aligned}\end{equation}
where $\epsilon_{\min}\coloneqq \min_k \epsilon_k$.
This evaluates the lower bound of \eqref{eq:multiple} as 

\begin{equation}\begin{aligned}
 \Delta e_{\max}&\geq\max_{\psi_{\rm in},\phi_{\rm in}}\frac{D_{\tr}(\psi_{\rm in},\phi_{\rm in})-2b_{\max}}{\sqrt{2\ln 2}\sqrt{Qn}\,K}\left(\frac{1}{1-\epsilon_{\min}}\right)^{L}.
\end{aligned}\end{equation}
Noting $\max_{\psi_{\rm in},\phi_{\rm in}}D_{\tr}(\psi_{\rm in},\phi_{\rm in})=1$ concludes the proof.
\end{proof}

\begin{remark}\label{remark:fidelity bound exponential}
Following a similar argument, we can prove that the fidelity-based lower bound \eqref{eq:fidelity bound} in the main text also shows the exponential growth with the number of layers. To see this, note that $1-\sqrt{F(\rho,\sigma)}\leq D_{\rm tr}(\rho,\sigma)$ holds for arbitrary states $\rho$ and $\sigma$. This gives
\bal
 \sqrt{1-F(\rho,\sigma)} &\leq \sqrt{1-(1-D_{\rm tr}(\rho,\sigma))^2}\\
 &=\sqrt{2D_{\rm tr}(\rho,\sigma)-D_{\rm tr}(\rho,\sigma)^2}\\
 &\leq\sqrt{2D_{\rm tr}(\rho,\sigma)}.
 \label{eq:lower bound of trace distance with fidelity}
\eal
Therefore, the denominator of the lower bound in \eqref{eq:fidelity bound} in the main text is bounded as 
\bal
\sqrt{1-\prod_{q=1}^Q\prod_{k=1}^K F\left(\mE_q^{(k)}(\psi),\mE_q^{(k)}(\phi)\right)}&=\sqrt{1- F\left(\otimes_{q=1}^Q\otimes_{k=1}^K\mE_q^{(k)}(\psi),\otimes_{q=1}^Q\otimes_{k=1}^K\mE_q^{(k)}(\phi)\right)}\\
&\leq \sqrt{2D_{\rm tr}\left(\otimes_{q=1}^Q\otimes_{k=1}^K\mE_q^{(k)}(\psi),\otimes_{q=1}^Q\otimes_{k=1}^K\mE_q^{(k)}(\phi)\right)}\\
&\leq \sqrt{2}\left[\sqrt{2\ln 2} \sqrt{Qn} K\right]^{1/2} (1-\epsilon_{\min})^{L/2}
\eal
where we used the multiplicativity of the fidelity under tensor product in the first line, Eq.~\eqref{eq:lower bound of trace distance with fidelity} in the second line, and Eq.~\eqref{eq:upper bound of trace distance exponential} in the third line.
We thus get that the fidelity-based lower bound is lower bounded by 
\bal
\max_{\psi_{\rm in},\phi_{\rm in}}\frac{D_{\tr}(\psi_{\rm in},\phi_{\rm in})-2b_{\max}}{\sqrt{2}\left[\sqrt{2\ln 2}\sqrt{Qn}\,K\right]^{1/2}}\left(\frac{1}{1-\epsilon_{\min}}\right)^{L/2},
\eal
which still grows exponentially with the layer number $L$.
\end{remark}


\section{Details for protocol benchmarking (Figs.~\ref{fig:local_dephasing_all},~\ref{fig:global_dep_all})}\label{app:protocol benchmarking}

Here, we evaluate the maximum spreads for error-mitigation strategies that appear in Figs.~\ref{fig:local_dephasing_all},~\ref{fig:global_dep_all} in the main text and their corresponding strategy-independent lower bounds for the local dephasing noise and the global depolarizing noise acting on an $n$-qubit system.

{\bf Probabilistic error cancellation} --- We first consider the local dephasing noise $\mZ_\epsilon^{\otimes n}$. 
The maximum spread for probabilistic error cancellation is known to be the diamond norm of the inverse map of the noise channel~\cite{Regula2021operational}. 
We thus get 
\bal
 \Delta e_{\max}^{\rm PEC} &= \|\left(\mZ_\epsilon^{\otimes n}\right)^{-1}\|_\diamond\\
 &=\|\mZ_\epsilon^{-1}\|_\diamond^n\\
 &=\frac{1}{(1-2\epsilon)^n}
\eal
where in the second line we used the multiplicativity of the diamond norm under tensor product, and the third line is due to the results in Refs.~\cite{takagi2020optimal,Regula2021operational}, which gives \eqref{eq:PEC local dephasing achievable} in the main text.
On the other hand, the lower bound can be evaluated as 
\bal
 \max_{\psi,\phi}\frac{D_{\rm tr}(\psi,\phi)}{D_{\rm tr}(\mZ_\epsilon^{\otimes n}(\psi),\mZ_\epsilon^{\otimes n}(\phi))}&\geq \frac{D_{\rm tr}(\GHZ_+,\GHZ_-)}{D_{\rm tr}(\mZ_\epsilon^{\otimes n}(\GHZ_+),\mZ_\epsilon^{\otimes n}(\GHZ_-))}
\eal
where $\GHZ_\pm=\dm{\GHZ_\pm}$ with $\ket{\GHZ_\pm}\coloneqq \frac{1}{\sqrt{2}}(\ket{0}^{\otimes n}\pm \ket{1}^{\otimes n})$ are $n$-qubit $\GHZ$ states. 
$D_{\rm tr}(\GHZ_+,\GHZ_-)=1$ as they are orthogonal. Also, $\GHZ_\pm$ are invariant under an even number of the application of $Z$, e.g., $Z_1\otimes Z_2\ket{\GHZ_\pm}=\ket{\GHZ_\pm}$, while it is transformed to the other GHZ state under an odd number of the application of $Z$, e.g., $Z_1\ket{\GHZ_\pm}=\ket{\GHZ_\mp}$. 
Therefore, 
\bal
 \mZ_\epsilon^{\otimes n}(\GHZ_\pm) &= \sum_{k=0}^{\lfloor n/2 \rfloor} \binom{n}{2k}(1-\epsilon)^{n-2k}\epsilon^{2k} \GHZ_\pm\\
 &\quad+\sum_{k=1}^{\lfloor (n+1)/2 \rfloor} \binom{n}{2k-1}(1-\epsilon)^{n-2k+1}\epsilon^{2k-1} \GHZ_\mp\\
 &=\frac{1+(1-2\epsilon)^n}{2}\GHZ_\pm + \frac{1-(1-2\epsilon)^n}{2}\GHZ_\mp,
 \label{eq:GHZ local dephasing}
\eal 
where we used 
\bal
\sum_{k=0}^{\lfloor n/2 \rfloor} \binom{n}{2k}(1-\epsilon)^{n-2k}\epsilon^{2k}&=\sum_{k=0}^{n} \binom{n}{k}(1-\epsilon)^{n-k}\frac{\epsilon^k+(-\epsilon)^{k}}{2}\\
&=\frac{1+(1-2\epsilon)^n}{2}.
\eal
\bal
\sum_{k=1}^{\lfloor (n+1)/2 \rfloor} \binom{n}{2k-1}(1-\epsilon)^{n-2k+1}\epsilon^{2k-1}&=\sum_{k=0}^{n} \binom{n}{k}(1-\epsilon)^{n-k}\frac{\epsilon^k-(-\epsilon)^{k}}{2}\\
&=\frac{1-(1-2\epsilon)^n}{2}.
\eal
This leads to 
\begin{equation}\begin{aligned}
&D_{\rm tr}(\mZ_\epsilon^{\otimes n}(\GHZ_+),\mZ_\epsilon^{\otimes n}(\GHZ_-))\\
&=(1-2\epsilon)^n\frac{1}{2}\|\GHZ_+ - \GHZ_-\|_1\\
&=(1-2\epsilon)^n,
\end{aligned}\label{eq:trace-distance-local-dephasing}\end{equation}
leading to the lower bound $1/(1-2\epsilon)^n$ in \eqref{eq:PEC local dephasing lower bound} in the main text.

As for the global depolarizing noise $\mD_\epsilon^{2^n}(\rho)=(1-\epsilon)\rho+\epsilon\mbI/2^n$, the maximum spread achievable by probabilistic error cancellation is given by $\Delta e_{\max}^{\rm PEC}=\frac{1+(1-2^{-2n+1})\epsilon}{1-\epsilon}$~\cite{takagi2020optimal,Jiang2020physical,Regula2021operational}.
The lower bound \eqref{eq:PEC global depolarizing lower bound} in the main text can be obtained by noting that for arbitrary states $\psi$ and $\phi$,
\bal
 \frac{D_{\rm tr}(\psi,\phi)}{D_{\rm tr}(\mD_\epsilon^{2^n}(\psi),\mD_\epsilon^{2^n}(\phi))} &= \frac{\frac{1}{2}\|\psi-\phi\|_1}{\frac{1}{2}\|(1-\epsilon)(\psi-\phi)\|_1}\\ 
 &=\frac{1}{1-\epsilon}.
\eal

{\bf Virtual distillation} --- Here, we consider $Q$-copy virtual distillation. We investigate the bound \eqref{eq:spread fixed observable} in Methods, which represents a fine-grained version of Theorem~\ref{thm:spread} that allows us to consider a specific observable of interest. 
Here, we consider the observable $A=\frac{1}{2}\otimes_{i=1}^n X_i$ and take $\psi=\GHZ_+$, $\phi=\GHZ_-$ as reference states.
We used $Q=2$ to plot Figs.~\ref{fig:local_dephasing_all}~and~\ref{fig:global_dep_all}. 
However, the following discussion holds for an arbitrary $Q$.

Let us first consider the local dephasing noise.
Since the distorted states are $\mZ^{\otimes n}(\GHZ_\pm)=\alpha \GHZ_\pm+ (1-\alpha)\GHZ_\mp$ where $\alpha\coloneqq\frac{1+(1-2\epsilon)^n}{2}$ as in \eqref{eq:GHZ local dephasing}, the form in \eqref{eq:virtual distillaton output} in Methods is identified as $\lambda^\psi=\lambda^\phi=\alpha$, $\lambda_2^\psi=\lambda_2^\phi=1-\alpha$, $\psi_2=\GHZ_-$, and $\phi_2=\GHZ_+$.
Then, using \eqref{eq:virtual estimator} in Methods, we get the achievable maximum spread 
\bal
 \Delta e_A &= 2\times\frac{1}{2}\times\alpha^{-Q}\\ &=\left(\frac{2}{1+(1-2\epsilon)^n}\right)^{Q}.
 \label{eq:achievable spread VD local dephasing}
\eal
To get a lower bound, the denominator of the right hand side of \eqref{eq:spread fixed observable} in Methods is evaluated as 
\bal
 D_{\rm LM}\left(\tilde\psi_Q^{(K)},\tilde\phi_Q^{(K)}\right)&\leq D_{\rm tr}\left(\tilde\psi_Q^{(K)},\tilde\phi_Q^{(K)}\right) \\
&=D_{\rm tr}\left(\left[\mZ_\epsilon^{\otimes n}(\GHZ_+)\right]^{\otimes Q},\left[\mZ_\epsilon^{\otimes n}(\GHZ_-)\right]^{\otimes Q}\right).
\label{eq:trace distance upper bound VD local dephasing}
\eal
Since the terms in the expansion of $\left[\mZ_\epsilon^{\otimes n}(\GHZ_\pm)\right]^{\otimes Q}=\left[\alpha\GHZ_\pm+(1-\alpha)\GHZ_\mp\right]^{\otimes Q}$ are orthogonal to each other, and there are $\binom{Q}{k}$ terms in the expansion that are tensor products of $k$ $\GHZ_\pm$'s and $Q-k$ $\GHZ_\mp$'s having the coefficient $\alpha^{Q-k}(1-\alpha)^k$, we can further get
\begin{equation}\begin{aligned}
&D_{\rm tr}\left(\left[\mZ_\epsilon^{\otimes n}(\GHZ_+)\right]^{\otimes Q},\left[\mZ_\epsilon^{\otimes n}(\GHZ_-)\right]^{\otimes Q}\right)\\
&= \frac{1}{2}\sum_{k=0}^{Q} \binom{Q}{k}\, \left|\alpha^{Q-k}(1-\alpha)^k-(1-\alpha)^{Q-k}\alpha^k\right|.
\label{eq:virtual local dephasing trace distance}
\end{aligned}\end{equation}
The first term of the numerator of the right-hand side of \eqref{eq:spread fixed observable} in Methods is 
\begin{equation}\begin{aligned}
&\Tr[(A+\mbI/2)(\dm{\GHZ_+}-\dm{\GHZ_-})]\\
&= \Tr\left[\frac{\otimes_{i=1}^n X_i+\mbI}{2}(\dm{\GHZ_+}-\dm{\GHZ_-})\right]\\
&=1.
\label{eq:virtual numerator first}
\end{aligned}\end{equation}
Using \eqref{eq:virtual bias} in Methods, bias can be computed as 
\bal
b_A(\psi)-b_A(\phi) &=  \left(\frac{\lambda_2}{\lambda}\right)^Q\Tr[A\psi_2]- \left(\frac{\lambda_2}{\lambda}\right)^Q\Tr[A\phi_2]\\
&=-\left(\frac{1-(1-2\epsilon)^n}{1+(1-2\epsilon)^n}\right)^Q.
\label{eq:virtual local dephasing bias}
\eal
Combining \eqref{eq:trace distance upper bound VD local dephasing}--\eqref{eq:virtual local dephasing bias} gives a lower bound 
\bal
 \frac{1-\left(\frac{1-(1-2\epsilon)^n}{1+(1-2\epsilon)^n}\right)^Q}{\frac{1}{2}\sum_{k=0}^{Q} \binom{Q}{k}\, \left|\alpha^{Q-k}(1-\alpha)^k-(1-\alpha)^{Q-k}\alpha^k\right|}.
\eal
In particular, when $Q=2$, this expression reduces to $\alpha^{-2}$, which coincides with the achievable spread $\Delta e_A$ in \eqref{eq:achievable spread VD local dephasing}.

Let us next consider the global depolarizing noise.
Let $\{\GHZ_t\}_{t=1}^{2^n}$ be a set of orthogonal states with $\GHZ_1\coloneqq\GHZ_+$ and $\GHZ_2\coloneqq\GHZ_-$.
Then, the distorted states are written as
\bal
\mD_\epsilon^{2^n}(\GHZ_+)&=(1-\epsilon) \GHZ_++ \epsilon \mbI/2^n\\
&= (1-(1-1/2^n)\epsilon)\GHZ_+ + \frac{\epsilon}{2^n} \sum_{t\neq 1} \GHZ_t\\
\mD_\epsilon^{2^n}(\GHZ_-)&=(1-\epsilon) \GHZ_-+ \epsilon \mbI/2^n\\
&= (1-(1-1/2^n)\epsilon)\GHZ_- + \frac{\epsilon}{2^n} \sum_{t\neq 2} \GHZ_t.
\label{eq:GHZ global depolarizing}
\eal
Therefore, the form in \eqref{eq:virtual distillaton output} in Methods is identified as $\lambda^\psi=\lambda^\phi=1-(1-1/2^n)\epsilon$, $\lambda_k^\psi=\lambda_k^\phi=\epsilon/2^n,\,k=2,\dots,2^n$, $\psi_2=\GHZ_-$,
$\phi_2=\GHZ_+$, and $\psi_k=\phi_k=\GHZ_k,\,k=3,\dots,2^n$.
The achievable spread is then obtained as 
\bal
\Delta e_A&=\frac{1}{\left[1-(1-1/2^n)\epsilon\right]^Q}.
\eal

To get a lower bound, the first term of the numerator of the right-hand side of \eqref{eq:spread fixed observable} in Methods is 1 as in \eqref{eq:virtual numerator first}.
Noting that the terms in the expansion of $\left[\mD_\epsilon^{2^n}(\GHZ_\pm)\right]^{\otimes Q}=\left[(1-\epsilon)\GHZ_\pm+\epsilon\mbI/2^n\right]^{\otimes Q}$ have the coefficient $\alpha^{Q-k}(1-\alpha)^k$ if they contain $k$ $\mbI/2^n$'s and there are $\binom{Q}{k}$ such terms, we have
\begin{equation}\begin{aligned}
&D_{\rm tr}\left(\left[\mD_\epsilon^{2^n}(\GHZ_+)\right]^{\otimes Q},\left[\mD_\epsilon^{2^n}(\GHZ_-)\right]^{\otimes Q}\right)\\
&= \frac{1}{2}\|(1-\epsilon)^{Q}(\GHZ_+^{\otimes Q}-\GHZ_-^{\otimes Q})+(1-\epsilon)^{Q-1}\epsilon\,(\GHZ_+^{\otimes (Q-1)}-\GHZ_-^{\otimes (Q-1)})\otimes \mbI/2\\
&\quad + \dots\|_1\\
&\leq \sum_{k=0}^{Q-1} \binom{Q}{k} (1-\epsilon)^{Q-k}\,\epsilon^k\\
&=1-\epsilon^Q.
\label{eq:virtual global dep trace distance}
\end{aligned}\end{equation}
Bias can also be computed as 
\bal
b_A(\psi)-b_A(\phi) &= \sum_{k=2}^{2^n} \left(\frac{\lambda_k}{\lambda}\right)^Q\Tr[A\psi_k]- \sum_{k=2}^{2^n} \left(\frac{\lambda_k}{\lambda}\right)^Q\Tr[A\phi_k]\\
&= \left(\frac{\lambda_2}{\lambda}\right)^Q\left(\Tr[A\GHZ_-]- \Tr[A\GHZ_+]\right)\\
&= -  \left(\frac{\epsilon/2^n}{1-(1-1/2^n)\epsilon}\right)^Q.
\label{eq:bias VD global depolarizing}
\eal
Eqs.~\eqref{eq:virtual numerator first}, \eqref{eq:virtual global dep trace distance}, and \eqref{eq:bias VD global depolarizing} give the lower bound
\bal
 \frac{1-\left(\frac{\epsilon/2^n}{1-(1-1/2^n)\epsilon}\right)^Q}{1-\epsilon^Q}.
 \label{eq:lower bound VD global depolarizing}
\eal

{\bf Extrapolation} --- 
We finally consider $R^{\mathrm{th}}$ order noise extrapolation. We again study the bound \eqref{eq:spread fixed observable} in Methods and consider the observable $A=\frac{1}{2}\otimes_{i=1}^n X_i$ and reference states $\psi=\GHZ_+$, $\phi=\GHZ_-$.
We first remark that since $\Delta e_A=\sum_{r=0}^R |\gamma_r|$ as in \eqref{eq:extrapolation spread formula} in Methods, the spread is fully determined by the choice of noise-boosting parameters $\{c_r\}_{r=0}^R$ in \eqref{eq:extrapolation condition} in Methods. 
Since a larger bias can admit a smaller spread, a natural way of choosing $\{c_r\}_{r=0}^R$ is to introduce a threshold bias $b_{\rm th}$ and take $\{c_r\}_{r=0}^R$ that minimizes $\Delta e_A$ under the condition that the bias for the reference states is smaller than the threshold bias, i.e., $|b_A(\GHZ_\pm)|\leq b_{\rm th}$.
To plot data points in Figs.~\ref{fig:local_dephasing_all}~and~\ref{fig:global_dep_all}, we ran this optimization at each noise strength $\epsilon$ with a fixed threshold bias $b_{\rm th}$ for each noise model.

Let us first consider the local dephasing noise. To compute bias, note that the expression in \eqref{eq:GHZ local dephasing} gives
\bal
 \Tr[A \mZ_\epsilon^{\otimes n}(\GHZ_+)] &= \frac{1}{2}\frac{1+(1-2\epsilon)^n}{2}- \frac{1}{2}\frac{1-(1-2\epsilon)^n}{2} \\
 &= \frac{(1-2\epsilon)^n}{2}\\
 \Tr[A \mZ_\epsilon^{\otimes n}(\GHZ_-)]&= -\frac{1}{2}\frac{1+(1-2\epsilon)^n}{2}+ \frac{1}{2}\frac{1-(1-2\epsilon)^n}{2} \\
 &= -\frac{(1-2\epsilon)^n}{2}.
\eal
We then get bias for a given set of $\{c_r\}_{r=0}^R$ as
\bal
 b_A(\GHZ_+) &= \sum_{r=0}^R \gamma_r\Tr[A\mZ_{c_r\epsilon}^{\otimes n}(\GHZ_+)]-\Tr(A\GHZ_+)\\
 &=\sum_{r=0}^R \frac{\gamma_r(1-2c_r\epsilon)^n}{2}-\frac{1}{2}\\
 b_A(\GHZ_-) &= \sum_{r=0}^R \gamma_r\Tr[A\mZ_{c_r\epsilon}^{\otimes n}(\GHZ_-)]-\Tr(A\GHZ_-)\\
 &=-\sum_{r=0}^R \frac{\gamma_r(1-2c_r\epsilon)^n}{2}+\frac{1}{2},
 \label{eq:extrapolation local dephasing bias}
\eal
resulting in $b_A(\GHZ_+)-b_A(\GHZ_-)=-1+\gamma_r(1-2c_r\epsilon)^n$.
By combining \eqref{eq:extrapolation local dephasing bias} and the trace distance between $\mZ_\epsilon^{\otimes n}(\GHZ_\pm)^{\otimes R+1}$ computed by  \eqref{eq:virtual local dephasing trace distance}, the corresponding lower bound is obtained as
\bal
\frac{\sum_{r=0}^R \gamma_r(1-2c_r\epsilon)^n}{\frac{1}{2}\sum_{k=0}^{R+1} \binom{R+1}{k}\, \left|\alpha^{R+1-k}(1-\alpha)^k-(1-\alpha)^{R+1-k}\alpha^k\right|}
\eal
with $\alpha=\frac{1+(1-2\epsilon)^n}{2}$.
Fig.~\ref{fig:local_dephasing_all} was plotted by setting $b_{\rm th}=0.2$ and $R=11$.
This threshold bias $b_{\rm th}$ was chosen to make sure that the threshold bias can be achieved by extrapolation with a sufficiently large order. 
We present the plot for $R=11$ because we did not see significant change in plots for $R>11$.

We next consider the global depolarizing noise.
To compute bias, note that
\bal
 \Tr[A \mD_\epsilon^{2^n}(\GHZ_+)] &= \frac{1-\epsilon}{2}\\
 \Tr[A \mD_\epsilon^{2^n}(\GHZ_-)]&=-\frac{1-\epsilon}{2}
 \label{eq:noisy expectation values global depolarizing}
\eal
We then get the bias for a given set of $\{c_r\}_{r=0}^R$ as
\bal
 b_A(\GHZ_+) &= \sum_{r=0}^R \gamma_r\Tr[A\mD_{c_r\epsilon}^{2^n}(\GHZ_+)]-\Tr(A\GHZ_+)\\
 &=\sum_{r=0}^R \gamma_r\frac{1-c_r\epsilon}{2}-\frac{1}{2}\\
 &=-\sum_{r=0}^R \frac{\gamma_rc_r\epsilon}{2}\\
 b_A(\GHZ_-) &= \sum_{r=0}^R \gamma_r\Tr[A\mD_{c_r\epsilon}^{\otimes n}(\GHZ_-)]-\Tr(A\GHZ_-)\\
 &=-\sum_{r=0}^R \gamma_r\frac{1-c_r\epsilon}{2}+\frac{1}{2}\\
 &=\sum_{r=0}^R \frac{\gamma_rc_r\epsilon}{2},
 \label{eq:extrapolation global dep bias}
\eal
giving $b_A(\GHZ_+)-b_A(\GHZ_-)=-\sum_{r=0}^R \gamma_r c_r \epsilon$.
By combining \eqref{eq:extrapolation global dep bias} and the trace distance between $\mD_\epsilon^{2^n}(\GHZ_\pm)^{\otimes R+1}$ computed by \eqref{eq:virtual global dep trace distance}, we get the lower bound as 
\bal
 \frac{1-\sum_{r=0}^R \gamma_r c_r \epsilon}{1-\epsilon^{R+1}}.
 \label{eq:lower bound EX global depolarizing}
\eal

Fig.~\ref{fig:global_dep_all} was plotted by setting $b_{\rm th}=0.1$ and $R=1$.
We chose these values because the noisy expectation values \eqref{eq:noisy expectation values global depolarizing} are linear in $\epsilon$ and thus the first order extrapolation can already realize the zero bias. 
(Therefore, the choice of $b_{\rm th}$ is rather arbitrary.)
Indeed, solving \eqref{eq:extrapolation condition} in Methods with $R=1$, we get $\gamma_0=c_1/(c_1-1)$ and $\gamma_1=-1/(c_1-1)$, which realizes $\gamma_0c_0+\gamma_1c_1=0$ noting $c_0\coloneqq 1$.
This observation allows us to further simplify \eqref{eq:lower bound EX global depolarizing} to $1/(1-\epsilon^2)$. 
This explains why the lower bound for extrapolation appears to coincide with the lower bound for virtual distillation in Fig.~\ref{fig:global_dep_all}.
The lower bound for virtual distillation \eqref{eq:lower bound VD global depolarizing} has the same denominator $1-\epsilon^2$ (as we chose $Q=2$), while the numerator of \eqref{eq:lower bound VD global depolarizing} becomes very close to 1 at a large qubit number such as $n=50$.

In addition, the achievable maximum spread for extrapolation appears to match the achievable maximum spread for probabilistic error cancellation in Fig.~\ref{fig:global_dep_all}.
This can be explained as follows. Since $\Delta e_A= |\gamma_0|+|\gamma_1|=\frac{c_1+1}{c_1-1}$ as in \eqref{eq:extrapolation spread formula} in Methods, the maximum spread is a decreasing function of $c_1$. 
Since the bias is zero and thus is always smaller than the threshold bias, the optimal $c_1$ always takes the maximum allowed value $1/\epsilon$.
This gives $\Delta e_A=\frac{1+\epsilon}{1-\epsilon}$, which is very close to $\Delta e_{\max}^{\rm PEC}=\frac{1+(1-2^{-2n+1})\epsilon}{1-\epsilon}$ at a large qubit number $n$.

\section{Comparison between trace-distance and subfidelity bounds}\label{app:subfidelity bound}

In the main text, we presented two alternative bounds \eqref{eq:fidelity bound} and \eqref{eq:subfidelity bound} that may admit easier evaluation than \eqref{eq:spread bound} and \eqref{eq:spread bound2} based on distinguishability measures.
In Remark~\ref{remark:fidelity bound exponential} in Supplementary Note~\ref{app:layer}, we saw that the fidelity-based bound \eqref{eq:fidelity bound} still shows the exponential growth with the circuit depth of layered circuits, confirming the effectiveness of the bound \eqref{eq:fidelity bound} in a certain setting. 
Here, we investigate the bound \eqref{eq:subfidelity bound}
based on subfidelity by comparing it to the trace-distance based bound \eqref{eq:spread bound2} in the representative settings discussed in Supplementary Note~\ref{app:protocol benchmarking}.

{\bf Local dephasing noise} --- Here, we calculate the subfidelity between the GHZ states under local dephasing noise. 
Using \eqref{eq:GHZ local dephasing}, $\mathrm{Tr}[\mZ_\epsilon^{\otimes n}(\mathrm{GHZ}_+)\mZ_\epsilon^{\otimes n}(\mathrm{GHZ}_-)]$ is calculated as follows:
\begin{equation}
    \begin{split}
        &\mathrm{Tr}[\mZ_\epsilon^{\otimes n}(\mathrm{GHZ}_+)\mZ_\epsilon^{\otimes n}(\mathrm{GHZ}_-)]\\
        &=\mathrm{Tr}\left[\left\{\frac{1+(1-2\epsilon)^n}{2}\GHZ_+ + \frac{1-(1-2\epsilon)^n}{2}\GHZ_-\right\}\left\{\frac{1+(1-2\epsilon)^n}{2}\GHZ_- + \frac{1-(1-2\epsilon)^n}{2}\GHZ_+\right\},\right]\\
        &=\mathrm{Tr}\left[\frac{1-(1-2\epsilon)^{2n}}{4}\mathrm{GHZ}_++\frac{1-(1-2\epsilon)^{2n}}{4}\mathrm{GHZ}_-\right]\\
        &=\frac{1-(1-2\epsilon)^{2n}}{2}.
    \end{split}
\end{equation}
We also have
\begin{equation}
    \begin{split}
        &\mathrm{Tr}[\mZ_\epsilon^{\otimes n}(\mathrm{GHZ}_+)\mZ_\epsilon^{\otimes n}(\mathrm{GHZ}_-)\mZ_\epsilon^{\otimes n}(\mathrm{GHZ}_+)\mZ_\epsilon^{\otimes n}(\mathrm{GHZ}_-)]\\
        &=\mathrm{Tr}\left[\left\{\frac{1-(1-2\epsilon)^{2n}}{4}\mathrm{GHZ}_++\frac{1-(1-2\epsilon)^{2n}}{4}\mathrm{GHZ}_-\right\}\left\{\frac{1-(1-2\epsilon)^{2n}}{4}\mathrm{GHZ}_++\frac{1-(1-2\epsilon)^{2n}}{4}\mathrm{GHZ}_-\right\}\right]\\
        &=2\left\{\frac{1-(1-2\epsilon)^{2n}}{4}\right\}^2.
    \end{split}
\end{equation}
Thus, the subfidelity is
\begin{equation}
    \begin{split}
        E(\mZ_\epsilon^{\otimes n}(\mathrm{GHZ}_+),\mZ_\epsilon^{\otimes n}(\mathrm{GHZ}_-))&=\frac{1-(1-2\epsilon)^{2n}}{2}+\sqrt{2\left[\left\{\frac{1-(1-2\epsilon)^{2n}}{2}\right\}^2-2\left\{\frac{1-(1-2\epsilon)^{2n}}{4}\right\}^2\right]}\\
        &=1-(1-2\epsilon)^{2n}.\label{eq:local-dephaseing-subfidelity}
    \end{split}
\end{equation}
This gives the denominator of the subfidelity bound \eqref{eq:subfidelity bound} in the main text as
\bal
\sqrt{1-E(\mZ_\epsilon^{\otimes n}(\mathrm{GHZ}_+),\mZ_\epsilon^{\otimes n}(\mathrm{GHZ}_-))^{KQ}} = \sqrt{1-\left[1-(1-2\epsilon)^{2n}\right]^{KQ}}.
\label{eq:subfidelity distance local dephasing}
\eal
In particular, when $K=Q=1$, this reduces to $(1-2\epsilon)^n$, which coincide with the trace distance \eqref{eq:trace-distance-local-dephasing}.

In Supplementary Figure~\ref{fig:subfidelity_local_dephasing}, we plot the relation between the subfidelity-based distance \eqref{eq:subfidelity distance local dephasing} and the trace distance computed by \eqref{eq:virtual local dephasing trace distance} for different choices of $K$ and $Q$. 
We can observe that the subfidelity can give good estimates of trace distance under the local dephasing noise acting on a system with $n=50$.

\begin{figure}
    \centering
    \includegraphics[width=.7\columnwidth]{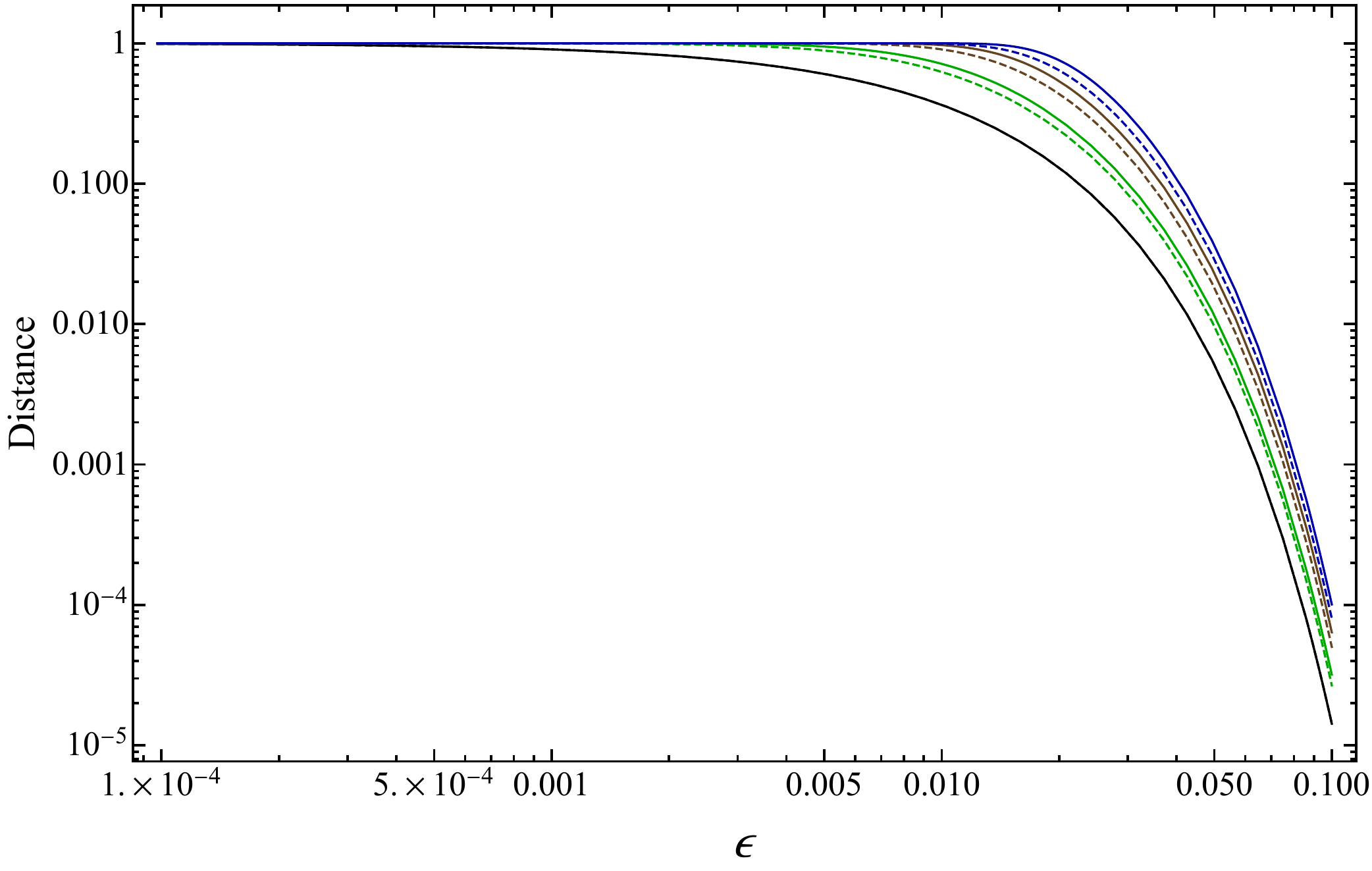}
    \caption{Subfidelity-based distance (solid) and trace distance (dahsed) under local dephasing noise on a $50$-qubit system with $QK=1$ (black), $5$ (green), $20$ (brown), and $50$ (blue). As explained in the text, the subfidelity-based distance and trace distance coincide for $QK=1$.}
    \label{fig:subfidelity_local_dephasing}
\end{figure}

{\bf Global depolarizing noise} --- Let us now consider the global depolarizing noise.
We first get 
\begin{equation}
    \begin{split}
        &\mathrm{Tr}[\mD_\epsilon^{2^n}(\mathrm{GHZ}_+)\mD_\epsilon^{2^n}(\mathrm{GHZ}_-)]\\
        &=\mathrm{Tr}\left[\left\{(1-\epsilon)\mathrm{GHZ}_++\frac{\epsilon\mathbb{I}}{2^n}\right\}\left\{(1-\epsilon)\mathrm{GHZ}_-+\frac{\epsilon\mathbb{I}}{2^n}\right\}\right]\\
        &=\mathrm{Tr}\left[\frac{\epsilon(1-\epsilon)}{2^n}\mathrm{GHZ}_++\frac{\epsilon(1-\epsilon)}{2^n}\mathrm{GHZ}_-+\frac{\epsilon^2\mathbb{I}}{2^{2n}}\right]\\
        &=\frac{\epsilon}{2^n}(2-\epsilon).
    \end{split}
\end{equation}
We also have
\begin{equation}
    \begin{split}
        &\mathrm{Tr}[\mD_\epsilon^{2^n}(\mathrm{GHZ}_+)\mD_\epsilon^{2^n}(\mathrm{GHZ}_-)\mD_\epsilon^{2^n}(\mathrm{GHZ}_+)\mD_\epsilon^{2^n}(\mathrm{GHZ}_-)]\\
        &=\mathrm{Tr}\left[\left\{\frac{\epsilon(1-\epsilon)}{2^n}\mathrm{GHZ}_++\frac{\epsilon(1-\epsilon)}{2^n}\mathrm{GHZ}_-+\frac{\epsilon^2\mathbb{I}}{2^{2n}}\right\}\left\{\frac{\epsilon(1-\epsilon)}{2^n}\mathrm{GHZ}_++\frac{\epsilon(1-\epsilon)}{2^n}\mathrm{GHZ}_-+\frac{\epsilon^2\mathbb{I}}{2^{2n}}\right\}\right]\\
        &=\left(\frac{\epsilon}{2^n}\right)^2\mathrm{Tr}\left[\left\{(1-\epsilon)\mathrm{GHZ}_++(1-\epsilon)\mathrm{GHZ}_-+\frac{\epsilon\mathbb{I}}{2^{n}}\right\}\left\{(1-\epsilon)\mathrm{GHZ}_++(1-\epsilon)\mathrm{GHZ}_-+\frac{\epsilon\mathbb{I}}{2^{n}}\right\}\right]\\
        &=\left(\frac{\epsilon}{2^n}\right)^2\mathrm{Tr}\left[\left\{(1-\epsilon)^2+\frac{(1-\epsilon)\epsilon}{2^{n-1}}\right\}\mathrm{GHZ}_++\left\{(1-\epsilon)^2+\frac{(1-\epsilon)\epsilon}{2^{n-1}}\right\}\mathrm{GHZ}_-+\frac{\epsilon^2\mathbb{I}}{2^{2n}}\right]\\
        &=\left(\frac{\epsilon}{2^n}\right)^2\left[2(1-\epsilon)^2+\frac{\epsilon(1-\epsilon)}{2^{n-2}}+\frac{\epsilon^2}{2^n}\right]
    \end{split}
\end{equation}
Therefore, the subfidelity is 
\begin{equation}
    \begin{split}
        E(\mD_\epsilon^{2^n}(\mathrm{GHZ}_+),\mD_\epsilon^{2^n}(\mathrm{GHZ}_-))&=\frac{\epsilon}{2^n}(2-\epsilon)+\left(\frac{\epsilon}{2^n}\right)\sqrt{2\left[2-\frac{\epsilon}{2^{n-2}}-\left(1-\frac{3}{2^n}\right)\epsilon^2\right]}.
    \end{split}
\end{equation}
This gives the denominator of the subfidelity bound \eqref{eq:subfidelity bound} in the main text as
\bal
\sqrt{1-E(\mD_\epsilon^{2^n}(\mathrm{GHZ}_+),\mD_\epsilon^{2^n}(\mathrm{GHZ}_-))^{KQ}}=\sqrt{1-\left\{\frac{\epsilon}{2^n}(2-\epsilon)+\left(\frac{\epsilon}{2^n}\right)\sqrt{2\left[2-\frac{\epsilon}{2^{n-2}}-\left(1-\frac{3}{2^n}\right)\epsilon^2\right]}\right\}^{KQ}}.
\label{eq:subfidelity distance global depolarizing}
\eal
When $n=K=Q=1$, this reduces to $1-\epsilon$, which coincides with the trace distance. 

To compare \eqref{eq:subfidelity distance global depolarizing} to the trace distance in more general cases, let us first get the exact expression of the trace distance with general $K$ and $Q$. 
Recall the second expression in \eqref{eq:GHZ global depolarizing}, 

\bal
\mD_\epsilon^{2^n}(\GHZ_+)&= (1-(1-1/2^n)\epsilon)\GHZ_+ + \frac{\epsilon}{2^n} \sum_{t\neq 1} \GHZ_t\\
\mD_\epsilon^{2^n}(\GHZ_-)&= (1-(1-1/2^n)\epsilon)\GHZ_- + \frac{\epsilon}{2^n} \sum_{t\neq 2} \GHZ_t.
\eal
Since all terms are orthogonal to each other, each term in the expansion of $\mD_\epsilon^{2^n}(\GHZ_\pm)^{\otimes QK}$ is also orthogonal. 
Also, the coefficient for each term of $\mD_\epsilon^{2^n}(\GHZ_+)^{\otimes QK}-\mD_\epsilon^{2^n}(\GHZ_-)^{\otimes QK}$ only depends on the number of $\GHZ_+$ and $\GHZ_-$ in the $QK$ subsystems. 
Namely, if a term in the expansion contains $k_1$ $\GHZ_+$ and $k_2$ $\GHZ_-$, then the coefficient for this term is 
\bal
\left[1-(1-1/2^n)\epsilon\right]^{k_1}(\epsilon/2^n)^{QK-k1} - \left[1-(1-1/2^n)\epsilon\right]^{k_2}(\epsilon/2^n)^{QK-k_2}.
\eal
Since there are $\binom{QK}{k_1}\binom{QK-k_1}{k_2}$ ways of choosing the location of $k_1$ $\GHZ_+$ and $k_2$ $\GHZ_-$, and $(2^n-2)^{QK-(k_1+k_2)}$  ways of choosing $\GHZ_i$ other than $\GHZ_\pm$ in remaining $QK-(k_1+k_2)$ subsystems (when $n\geq 2$), there are $\binom{QK}{k_1}\binom{QK-k_1}{k_2}(2^n-2)^{QK-(k_1+k_2)}$ such terms in the expansion. 
Thus, we get for $n\geq 2$,
\begin{equation}\begin{aligned}
&D_{\rm tr}\left(\left[\mD_\epsilon^{2^n}(\GHZ_+)\right]^{\otimes QK},\left[\mD_\epsilon^{2^n}(\GHZ_-)\right]^{\otimes QK}\right)\\
&= \frac{1}{2}\sum_{k_1=0}^{QK}\sum_{k_2=0}^{QK-k_1}\binom{QK}{k_1}\binom{QK-k_1}{k_2}(2^n-2)^{QK-(k_1+k_2)}\left|\left[1-\left(1-\frac{1}{2^n}\right)\epsilon\right]^{k_1}\left(\frac{\epsilon}{2^n}\right)^{QK-k_1}-\left[1-\left(1-\frac{1}{2^n}\right)\epsilon\right]^{k_2}\left(\frac{\epsilon}{2^n}\right)^{QK-k_2}\right|.
\label{eq:virtual global dep trace distance exact}
\end{aligned}\end{equation}

\begin{figure}
    \centering
    \subfloat[][]{\includegraphics[width=.45\columnwidth]{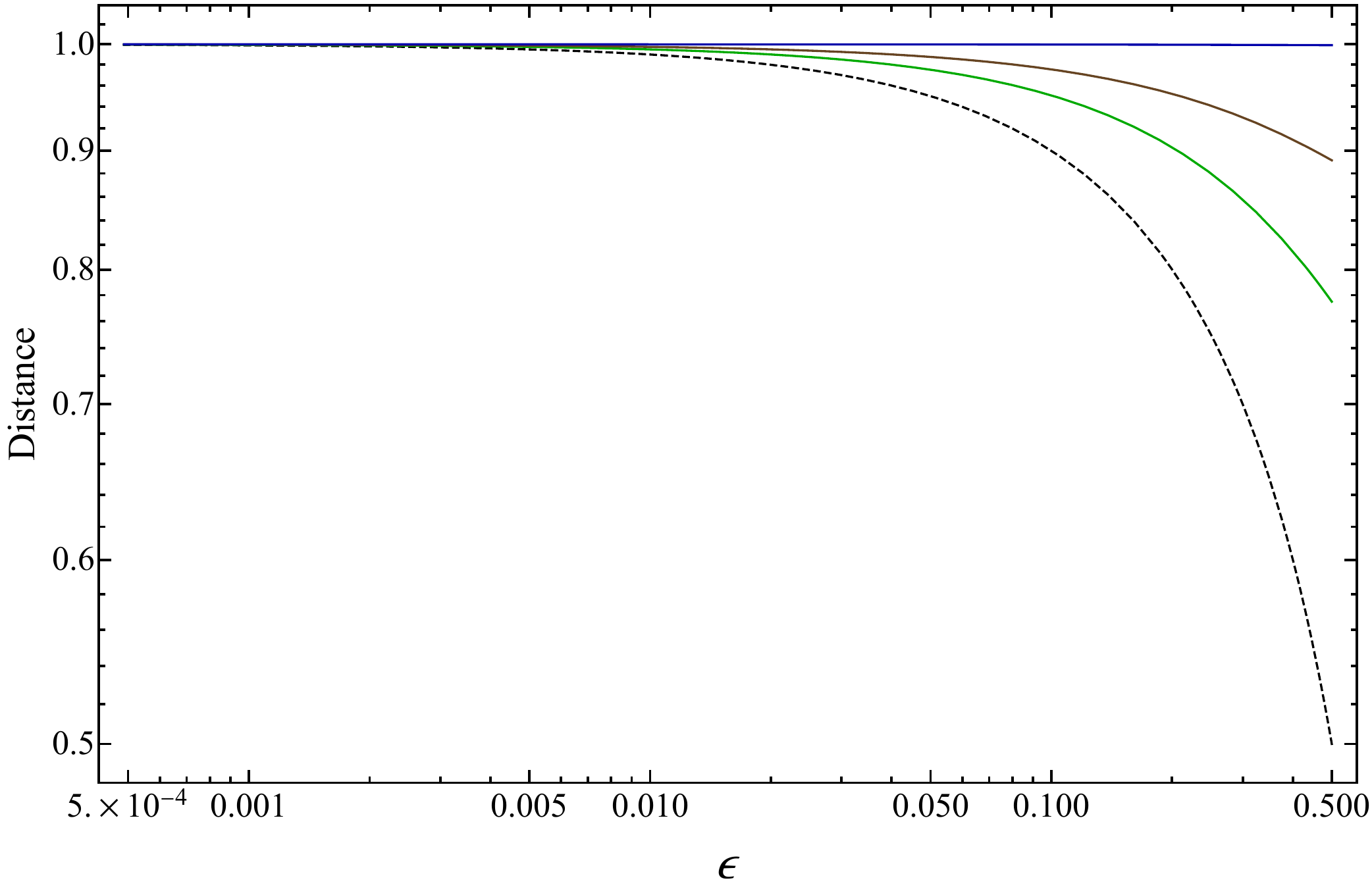}}\quad
    \subfloat[][]{\includegraphics[width=.45\columnwidth]{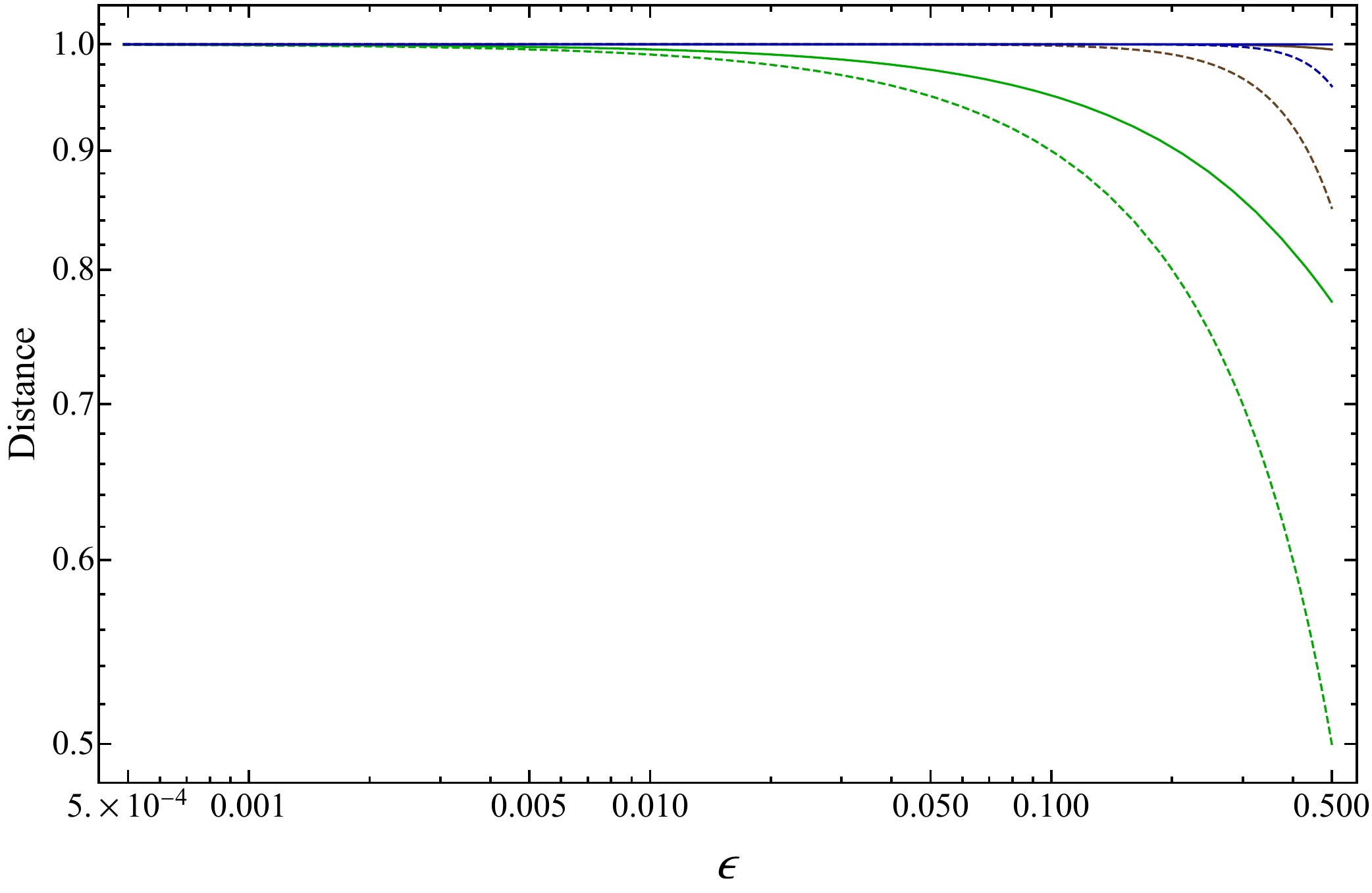}}
    \caption{Subfidelity-based distance and trace distance under global depolarizing noise. (a) $QK=1$ is fixed. In this case, the trace distance becomes independent of $n$, which is shown by a black dashed curve. Solid curves denote the subfidelity-based distance with $n=2$ (green), 3 (brown), and 10 (blue). (b) $n=2$ is fixed. The solid and dashed curves respectively denote subfidelity-based distance and trace distance with $QK=1$ (green), 5 (brown), and 10 (blue). Note that the brown and blue solid curves almost overlap.}
    \label{fig:subfidelity_global_depolarizing}
\end{figure}

Supplementary Figure~\ref{fig:subfidelity_global_depolarizing} shows the relation between the subfidelity-based distance and the trace distance for different choices of $n$, $K$, and $Q$. 
We can see that the subfidelity tends to give better estimates for the small error region.
In the larger error region, the subfidelity quickly becomes loose as $n$ increases as can be seen in Supplementary Figure~\ref{fig:subfidelity_global_depolarizing}\,(a).
On the other hand, we can see in Supplementary Figure~\ref{fig:subfidelity_global_depolarizing}\,(b) the tendency that the gap between the two distances becomes smaller as $K$ and $Q$ increase. 

\end{document}